\RequirePackage{fix-cm}
\documentclass[twocolumn]{svjour3}          % twocolumn
\smartqed  % flush right qed marks, e.g. at end of proof
\usepackage{graphicx}
\usepackage{mathptmx}      % use Times fonts if available on your TeX system
\usepackage{amsmath}
\usepackage{amssymb}
\usepackage{mathrsfs}
\usepackage{bbm}
\usepackage{xcolor}
\usepackage{array}
\usepackage{cite} 
\usepackage{hyperref}
\usepackage{abstract}
\usepackage{multicol}
\usepackage{multirow}
\usepackage{longtable}
\usepackage{threeparttablex}
\journalname{arXiv}
\begin{document}

%------------------------------------------------------------

\title{Mass evolution of Schwarzschild black holes%\thanks{Grants or other notes
%about the article that should go on the front page should be
%placed here. General acknowledgments should be placed at the end of the article.}
}
%\subtitle{Do you have a subtitle?\\ If so, write it here}

%\titlerunning{Short form of title}        % if too long for running head

\author{Natal\'i Soler Matubaro de Santi \and Raphael Santarelli}

%\authorrunning{Short form of author list} % if too long for running head

\institute{Natal\'i Soler Matubaro de Santi \at
              Physics Department, Federal University of S\~ao Carlos, SP, Brazil \\
              ORCID: \href{https://orcid.org/0000-0002-4728-6881}{ORCID: 0000-0002-4728-6881}\\
              \email{natalidesanti@gmail.com}           %  \\
           \and
           Raphael Santarelli \at
           Physics Department, Federal University of S\~ao Carlos, SP, Brazil \\
           ORCID: \href{https://orcid.org/0000-0002-3742-8153}{ORCID: 0000-0002-3742-8153}\\
           \email{santarelli@df.ufscar.br}
}
\date{\today} 
%\date{Received: date / Accepted: date}
% The correct dates will be entered by the editor

%------------------------------------------------------------

\maketitle

%------------------------------------------------------------

\begin{abstract}
 In the classical theory of general relativity black holes can only absorb and not emit particles. When quantum mechanical effects are taken into account, then the black holes emit particles as hot bodies with temperature proportional to $\kappa$, its surface gravity. This thermal emission can lead to a slow decrease in the mass of the black hole, and eventually to its disappearance, also called \textit{black hole evaporation}. This characteristic allows us to analyze what happens with the mass of the black hole when its temperature is increased or decreased, and how the energy is exchanged with the external environment. This paper has the aim to make a review about the mass evolution of Schwarzschild black holes with different initial masses and external conditions as the empty space, the cosmic microwave background with constant temperature, and with temperature varying in accordance with the eras of the universe. As a result, we have the complete evaporation of the black holes in most cases, although their masses can increase in some cases, and even diverge for specific conditions.

\end{abstract}

%------------------------------------------------------------

\section{Introduction}
\label{sec:intro}

The theory of General Relativity (GR) was published by Albert Einstein in 1915 \cite{einstein1915} and changed completely our description of the universe. Since now gravity is no longer an action-at-a-distance force, but a geometrical property of a curved space-time, i.e., the gravitational field can be interpreted as the result of the space-time curvature \cite{carroll2004, hartle2003, wald1984, schutz2009, dinverno1998, misner1973}. The black holes are one of the most interesting predictions of the theory, and they appear at the end stage of very massive stars collapse, when no internal process can generate the necessary pressure to balance the gravitational attraction \cite{chandrasekhar1935, penrose2002, oppenheimer1939}. These objects are characterized by singularity and an event horizon, from which classically nothing can escape, not even light \cite{brehme1977, wheeler1971, blandford1999, schwarzschild1916, kerr1963}.

Although GR has passed many experimental tests in the last century \cite{schiff1960, will2001} there are some reasons to suppose it is not the final theory for gravitation, because it explicitly predicts singularities (as the ones inside black holes) \cite{hawking1970, camacho2008, shomer2007, deser1957} and it does not agree with the Quantum Field Theory (QFT) in the appropriated limit. Even thought there is still no satisfactory candidate to unify GR and QFT, a minimally consistent junction is made by the Quantum Field Theory in Curved Space-time (QFTCS). In this context the matter fields are quantized, but gravity is still classically described by GR \cite{maldacena2009, mukhanov2007, parker2009, birrell1982, fulling1982, wald1994}.

In 1975 Stephen Hawking published his emblematic paper ``Particle Creation by Black Holes'' \cite{hawking1975}, where he studied effects of a quantum field in a classical background using the semi-classical approach of QFTCS. It caused scientific upheaval at that time because the result was opposed to the classical idea that black holes are passive objects, in sense that they just absorb matter and never emit them (and hence have zero temperature). By quantizing the scalar field in a spherical symmetric collapse scenario, Hawking proved that black holes emit particles asymptotically in the future, and therefore have non-zero temperature.

Even though the Hawking paper officially inaugurated the so called {\em black hole thermodynamics}, the idea that black holes should behave as thermodynamics objects appeared previously, in 1973 with Jacob Bekenstein \cite{bekenstein1973}. Based on the {\em Hawking area theorem} \cite{hawking1972}, that says that the area of the event horizon of a black hole cannot decrease, Bekenstein established a formal analogy between the laws of thermodynamics and the laws of black hole mechanics, defining an entropy for the black hole related with its area. This result showed that black holes physics is intrinsically {\em irreversible}, presenting an {\em arrow of time}, just like thermodynamics \cite{bekenstein1980, traschen1999, lambert2013, parker1980, lopresto2003, hawking1976, bekenstein1975, belgiorno2004}. However, the analogy proposed by Bekenstein was incomplete because it did not include the temperature of the black hole, only the entropy, since classically all black holes possess zero temperature, and he could not solve this inconsistency. It was Hawking's paper \cite{hawking1975} (using QFTCS) that gave the correct meaning to the temperature and entropy of black holes.

With the expressions for the temperature and entropy defined, the black holes can be analyzed as thermodynamics objects, and an interesting property of Schwarzschild black holes, that will be shown in the present paper, is that they have negative heat capacity, a feature that does not allow them to achieve thermal equilibrium with an energy reservoir. Thus, a Schwarzschild black hole in empty space will lose energy and increase its temperature until its mass approaches zero. When this happens we have the so called {\em black hole evaporation}, although the final configuration is still unknown \cite{unruh2017}. Since a black hole in empty space is an idealization, we will also study the mass evolution of Schwarzschild black holes when immersed in the cosmic microwave background radiation with constant temperature and with temperature evolving in time according to the universe eras: radiation, matter and dark energy \cite{carroll2004, hartle2003, thorne2017}.

The objective of this paper is briefly present the laws of black hole thermodynamics (see Section \ref{sec:TL}), then show that black holes cannot achieve a thermal equilibrium with an energy reservoir (see Section \ref{sec:inaccessibility}) and analyze the mass evolution of them (see Sections \ref{sec:empty}, \ref{sec:rad} and \ref{sec:ETU}). At the end we discussed our results.

%------------------------------------------------------------

\section{Analogy between black hole mechanics and thermodynamics}
\label{sec:TL}

\begin{table*}
 \caption{\label{tab:T1}Thermodynamic laws and Classical laws for black holes}
 \begin{center}
  \begin{tabular}{ccc}
   \cline{1-3}
   \textbf{\footnotesize Law} & \textbf{\footnotesize Thermodynamic laws} & \textbf{\footnotesize Black hole laws} \\
   \cline{1-3}
   \multirow{2}{*}{\footnotesize 0th} & \multicolumn{1}{c}{\footnotesize Equality of temperatures is a condition for thermodynamic} & \multicolumn{1}{c}{\footnotesize The surface gravity $\kappa$ of a stationary black}\\
    & \multicolumn{1}{c}{\footnotesize equilibrium between systems or between parts of the same system} & \multicolumn{1}{c}{\footnotesize  hole is constant over its entire horizon}\\
   \cline{1-3}
   \multirow{2}{*}{\footnotesize 1st} & \multicolumn{1}{c}{\footnotesize In an isolated system, the total energy of that system} & \multicolumn{1}{c}{\footnotesize In an isolated system, including black holes, the total energy}\\
   & {\footnotesize is conserved: $d U = T dS - d W$} & \multicolumn{1}{c}{\footnotesize is conserved: $d M = \frac{\kappa}{8 \pi} dA + \Omega_H dL + \Phi_H d Q$}\\
   \cline{1-3}
   \multirow{2}{*}{\footnotesize 2nd} & \multicolumn{1}{c}{\footnotesize In an isolated system, during any process, the total entropy of} & \multicolumn{1}{c}{\footnotesize The sum of the entropy of the black hole with its exterior} \\
   & \multicolumn{1}{c}{\footnotesize the system increases or remains the same: $\delta S_T \ge 0$} & \multicolumn{1}{c}{\footnotesize increases or remains the same: $\delta S_{B H} + \delta S_{ext} \ge 0$} \\
   \cline{1-3}
   \multirow{2}{*}{\footnotesize 3rd} & \multicolumn{1}{c}{\footnotesize It is impossible to reduce the temperature of a system to} & \multicolumn{1}{c}{\footnotesize It is impossible to reduce the surface gravity $\kappa$ of a black}\\
   & \multicolumn{1}{c}{\footnotesize zero by a finite number of processes} & \multicolumn{1}{c}{\footnotesize hole to zero by a finite number of processes}\\
   \cline{1-3}
   \end{tabular}
  \end{center}
\end{table*}

Accordingly with \cite{hawking1975}, the expression for the entropy $S$ and temperature $T$ of a Schwarzschild black hole are:
\begin{align}
  T & = \frac{\kappa}{2 \pi} = \frac{1}{8 \pi M} , \label{eq:SBHT}\\
  S & = \frac{A}{4} = 4 \pi M^2 ,
\end{align}
where $M$ is the mass, $A = 16 \pi M^2$ is the event horizon area and $\kappa = \frac{1}{4 M}$ is the surface gravity evaluated at the horizon of the black hole \cite{carroll2004, lambert2013, wald1984, hartle2003, mukhanov2007, lopresto2003, parker1980}. Also, the black hole mass $M$ is identified with its energy: $E = M$. Note that, in the above expressions (and in many other expressions in the present paper), we are going to use Planck units, i.e., we are considering the speed of light $c$, the universal gravitational constant $G$, the reduced Planck constant $\hbar$ and the Boltzmann constant $k_B$ all equal to 1. We will retake those constants when convenient.

With these explicit equations, we are now able to relate the laws of thermodynamics with the laws of black hole mechanics for a Schwarzschild black hole. This analogy can be seen in the table \ref{tab:T1}. The {\em 0th law} connects the temperature $T$ of the black hole with its surface gravity $\kappa$. The {\em 1th law} states that the mass of black hole $M$ represents its total energy, related with the thermodynamic internal energy $U$ of a system, and that the work term $d W$ is represented by the variation of angular momentum $d L$ and charge $d Q$, with $\Omega_H$ representing the angular speed and $\Phi_H$ the electric potential at the event horizon. Finally, the {\em 2th law} links the area $A$ of the horizon to the entropy $S$ of the black hole.

%------------------------------------------------------------

\section{The inaccessibility of thermal equilibrium with an energy reservoir and its implications}
\label{sec:inaccessibility}

In classical thermodynamics, bodies exchanging energy with a thermal reservoir eventually achieve a stable equilibrium, i.e., these bodies acquires the reservoir temperature in a finite time and remain with it. However, for a Schwarzschild black hole, this thermal equilibrium is never achieved. To show this, let's first derive the heat capacity of a Schwarzsch-ild black hole \cite{callen1985, reif1965}:
\begin{equation}
 C = \frac{d E}{d T} = \frac{d M}{d T} = - \frac{1}{8 \pi T^2} = - 8 \pi M^2 < 0 . \label{eq:CBN}
\end{equation}
Notice that $C < 0$, and the negative sign is responsible for this unusual feature. Let's interpret this result in terms of energy flux between a thermal reservoir and a black hole. If the black hole, with initial temperature $T$, is in contact with an energy reservoir with temperature $T_R \neq T$, we can have two different situations:
(i) If $T > T_R$, the black hole will emit radiation and transfer heat to the reservoir. Since its energy is related with its mass, the black hole will lose mass, in a process called \textit{evaporation}. And since its temperature is inversely proportional to its mass, the black hole temperature will increase, moving further away from the equilibrium.
(ii) If $T < T_R$, the black hole will receive energy, increase its mass and decrease its temperature, also moving further away from the equilibrium.
Therefore, in both cases, the black hole never achieves the reservoir temperature $T_R$. If initially $T = T_R$, there is no net flux, and the temperature of the black hole does not change. Consequently, its mass also remains constant.

%-----------------------------------------------------------------------

\section{Black hole and the empty space}
\label{sec:empty}

Consider a Schwarzschild black hole with initial mass $M_0$ and initial temperature $T_0 >0$ in empty space. Since this black hole is immersed in the vacuum, we can consider a reservoir with zero temperature, $T_R = 0$. The black hole will then emit radiation, lose its mass and increase its temperature, until the complete evaporation. In order to compute how the mass changes with time, we will use the Stefan-Boltzmann law for the power emitted by the black hole, and we will also \textit{assume} that the black hole emits as a perfect black body\footnote{We need to evidence that here we are making an approximation. We are using the result \ref{eq:SBHT}, for a Schwarzschild black hole and considering a perfect black body emission, using the Stefan-Boltzmann law. In order to take this more rigorously we should consider the angular dependency (from quantized field) and the back-scattering (from probability conservation) in the temperature derivation \cite{birrell1982, mukhanov2007, parker2009}. In this case we would be dealing with a black hole as a {\em gray body}. The reference \cite{page1976} considers them and estimates a numerical factor $\alpha = 10^{- 4}$ in equation \ref{eq:STvazio}, rewriting it as $L = \alpha \sigma A T^4$.} \cite{callen1985, reif1965}. The power emitted is then:
\begin{equation}
 L = \sigma A T^4 = \frac{\hbar c^6}{15360 \pi G^2} \frac{1}{M^2} , \label{eq:STvazio}
\end{equation}
where $\sigma = \frac{\pi^2 k_B^4}{60 \hbar^3 c^2}$ is Stefan-Boltzmann constant, $A = 4 \pi r_S^2$ is Schwarzschild black hole area with Schwarzschild radius $r_{S} = \frac{2 G M}{c^2}$ and temperature $T = \frac{\hbar c^3}{8 \pi G k_B M}$. Note that we are recalling the constants to use the international unit system (SI) in order to compare our results with physical masses and temperatures. Thus, the black hole mass decreases in time according to
\begin{equation}
 \frac{d \left(c^2 M\right)}{d t} = - L = - \frac{\hbar c^6}{15360 \pi G^2} \frac{1}{M^2} . \label{eq:Mt}
\end{equation}

\begin{figure}[h!]
  \centering
  \includegraphics[scale=0.33]{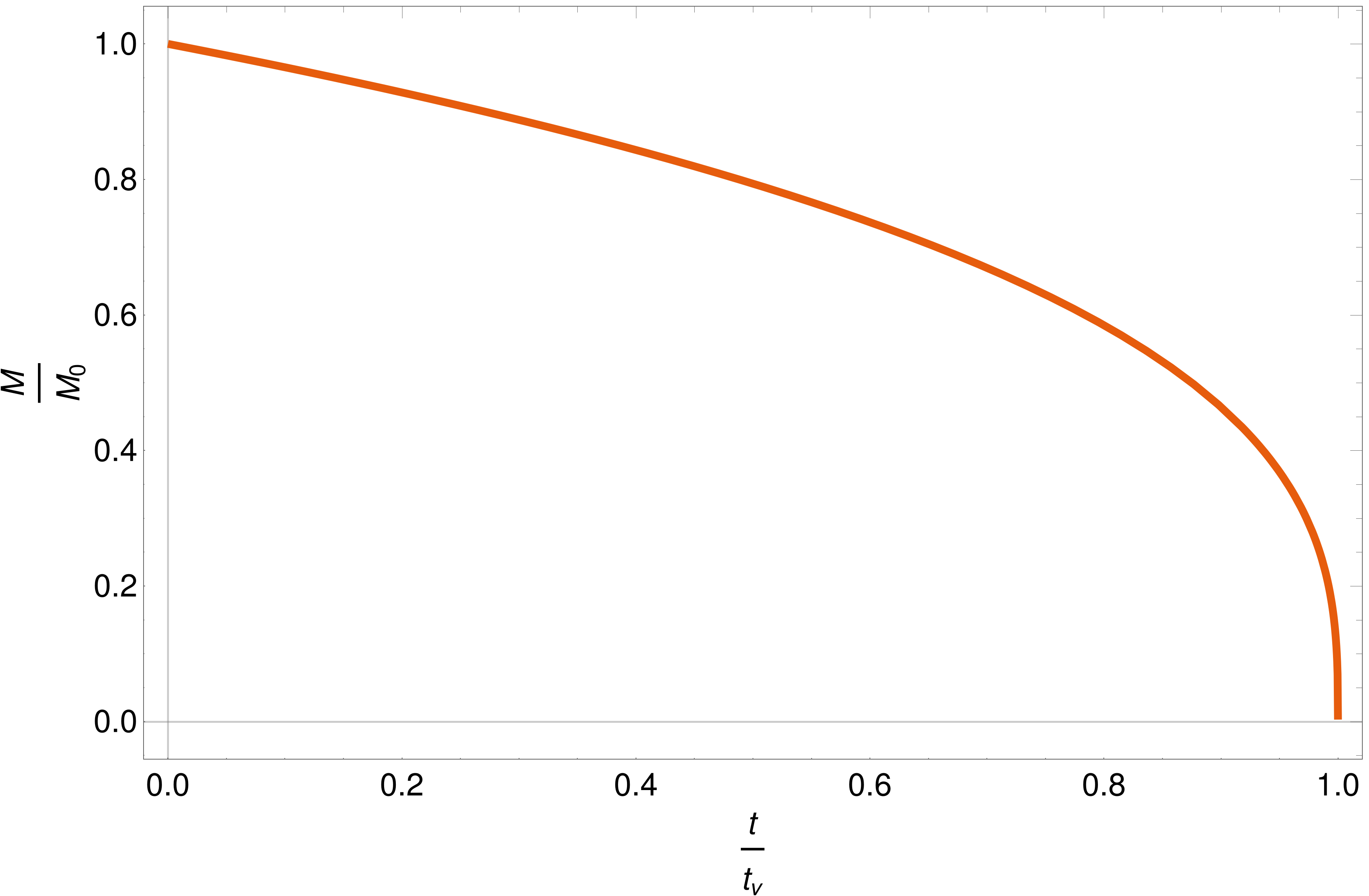}
  \caption{General mass decay of a Schwarzschild black hole in empty space. \label{fig:dec}}
\end{figure}

The equation (\ref{eq:Mt}) has a simple analytical solution. Considering the initial time $t = 0$, the integration constant is translated as initial mass of black hole $M (t = 0) = M_0$. Therefore we have
\begin{equation}
 \label{eq:M(t)}
 M (t) = M_0 \left(1 - \frac{t}{t_V}\right)^{1/3} , \hspace{0.5cm}t \in [0, t_V)
\end{equation}
where $t_V = 5120 \pi \frac{G^2}{\hbar c^4} M_0^3$ is the {\em lifetime}\footnote{By lifetime we refer to the necessary time until that black hole mass becomes zero, i.e., the necessary time until its complete evaporation.} of black hole. The general behavior of the mass decay of a Schwarzschild black hole can be viewed in the Figure \ref{fig:dec}, with intervals $\frac{M}{M_0}, \frac{t}{t_V} \in [0, 1)$.

For astronomical values of mass, the lifetime $t_V$ of a black hole is normally a huge number. Table \ref{tab:tv} shows this value for three different initial masses:
(i) the mass of Hyperion \cite{mhip},
(ii) the mass of the Sun \cite{pdg2014} and
(iii) the mass of the supermassive black hole in the center of Milky Way Galaxy \cite{massaVL}.

\begin{table}[h!]
\begin{center}
\caption{\label{tab:tv}Lifetime of black holes with determined masses.}
\begin{tabular}{c|cc} 
  \cline{2-3}
   & \textbf{\footnotesize M (kg)} & $\mathbf{\footnotesize t_V}$ \textbf{\footnotesize (s)} \\
   \cline{1-3}
   \multicolumn{1}{c|}{\textbf{\footnotesize Hyperion}}  & {\footnotesize $5.6 \cdot 10^{18}$} & {\footnotesize $1.5 \cdot 10^{40}$}\\
   \cline{1-3}
   \multicolumn{1}{c|}{\textbf{\footnotesize Sun}}  & {\footnotesize $2.0 \cdot 10^{30}$} & {\footnotesize $6.6 \cdot 10^{74}$}\\
   \cline{1-3}
   \multicolumn{1}{c|}{\multirow{1}{*}{\textbf{\footnotesize Black hole of}}} & {\footnotesize $4.5 \cdot 10^{6}$ $M_{\bigodot}$} & \multirow{2}{*}{\footnotesize{$6.0 \cdot 10^{94}$}} \\
   \multicolumn{1}{c|}{\multirow{1}{*}{\textbf{\footnotesize Milk Way center}}} & {\footnotesize $9.0 \cdot 10^{36}$} &\\
   \cline{1-3}
\end{tabular}
\end{center}
\end{table}

Notice that these values are much greater than the age of the universe $t_{uni} \simeq 4.4 \cdot 10^{17} s$ \cite{pdg2014}, which implies that, in practice, these black holes  will not evaporate in this time scale. For a black hole formed in the beginning of the universe to evaporate within the time scale of the universe, the initial mass must be inside the interval $M_0 \in (10^{- 9}, 10^{11}) kg$. These values were obtained using $t_{V} = t_{P} \simeq 5.4 \cdot 10^{- 44} s$ (Pla-nck time \cite{pdg2014, faraoni2017}) and $t_V = t_{uni}$ respectively.

%---------------------------------------------------------------------

\section{Black hole and the cosmic microwave background}
\label{sec:rad}

Now we are going to consider a Schwarzschild black hole with mass $M$ and temperature $T$ immersed in cosmic microwave background radiation (CMB). This radiation will act as a thermal reservoir to the black hole. In this first approach we will consider that the CMB have constant temperature $T_{CMB} \simeq 2.7255 K$ \cite{pdg2014}. There will be an exchange of energy between black hole and the CMB, and we can use again the Stefan-Boltzmann law as a thermal current between them. So, parallel to equation (\ref{eq:STvazio}), we have
\begin{equation}
 J = \sigma A \left(T^4 - T_{CMB}^4\right) .
\end{equation}
As we have made in equation (\ref{eq:Mt}), we can relate $J = - \frac{d (M c^2)}{d t}$, finding
\begin{equation}
 \label{eq:dJm} \frac{d M}{d t} = - \frac{a}{M^2} + b M^2 ,
\end{equation}
where $a \equiv \frac{\hbar c^4}{15360 \pi G^2} \simeq 4.0 \cdot 10^{15} \frac{kg^3}{s}$ and $b \equiv \frac{4 \pi^3 k_B^4 G^2 T_{CMB}^4}{15 \hbar^3 c^8} \simeq 9.6 \cdot 10^{- 76} \frac{1}{s \cdot kg}$. Here, we also have a complete analytical solution for the differential equation \ref{eq:dJm}:
\begin{align}
  t & = \frac{1}{4\hspace{0.05cm}b^{3/4}\hspace{0.05cm}a^{1/4}} \left\{ 2 \arctan \left[\left(\frac{b}{a}\right)^{1/4} M\right] \right. \nonumber\\
    & \left. \hspace{1cm} + \log \left(a^{1/4} - b^{1/4} M\right) - \log \left(a^{1/4} + b^{1/4} M\right)\right\} . \label{eq:dif1}
\end{align}
This equation is not so simple as equation \ref{eq:M(t)} but it expresses different behaviors for different ranges of mass $M$, including unphysical ones for negative masses when $M > - \left(\frac{a}{b}\right)^{1/4}$ (this will be discussed later). This complete analytical solution was used to calibrate the complete numerical one and to check the code used, since in the next section only the complete numerical solutions will be used (together with the asymptotic analytical solutions, because the complete differential equations will be more complicated). Moreover, according to initial temperature of the black hole, we have three different situations:
(i) the black hole will initially emit energy to CMB if $T > T_{CMB}$,
(ii) the black hole will initially absorb energy from CMB if $T < T_{CMB}$, and
(iii) there will be no net flux if $T = T_{CMB}$.
Therefore, the right side (RHS) of equation (\ref{eq:dJm}) can be negative, positive or zero, respectively.

Since the temperature of the reservoir does not change with time, we know that (accordingly with the discussion in section \ref{sec:inaccessibility}) if $T > T_{CMB}$, then the black hole temperature will always be greater than $T_{CMB}$; if $T < T_{CMB}$ the temperature will always be smaller than $T_{CMB}$; and if $T = T_{CMB}$ the temperature will be constant and equal to $T_{CMB}$. Hence, the radiation temperature $T_{CMB}$ separates the three different behaviors, and we can compute the mass of the black hole with temperature equivalent to $T_{CMB}$ as
\begin{equation}
 M_{CMB} = \frac{\hbar c^3}{8 \pi G k_B T_{CMB}} \simeq 4.5 \cdot 10^{22} kg . \label{eq:Mrcf}
\end{equation}
In terms of the mass, for initial mass of black hole $M = M_{CMB}$, the RHS of equation (\ref{eq:dJm}) will be zero, and the mass (and also the temperature) will be constant. For $M < M_{CMB}$, the RHS of equation (\ref{eq:dJm}) will be negative, and the mass will decrease with time, eventually evaporating (with the temperature diverging). Finally, for $M > M_{CMB}$, the RHS of equation (\ref{eq:dJm}) will be positive, and the mass will increase with time (as the temperature approaches zero).

\begin{figure}[h!]
\centering
\includegraphics[scale=0.32]{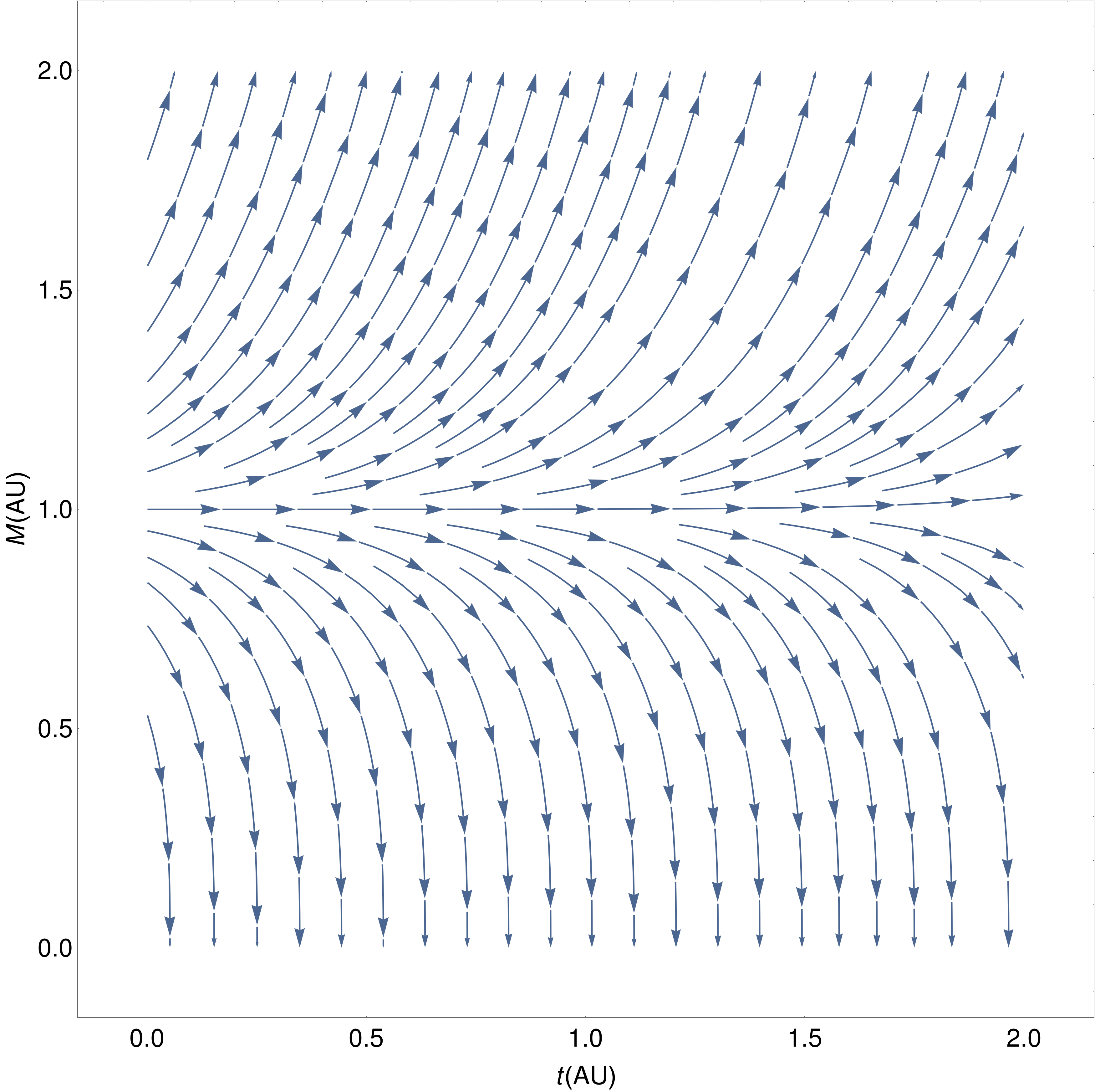}
\caption{\label{fig:RCRcte} Stream plot (mass {\em versus} time) for a Schwarzschild black hole immersed in cosmic microwave background radiation with constant temperature. Note that we used arbitrary units (AU).}
\end{figure}

Analyzing the differential equation (\ref{eq:dJm}) using \emph{stream plot} we obtained the Figure \ref{fig:RCRcte}, in arbitrary units. The goal of this figure is to show the general behavior of the solutions of equation (\ref{eq:dJm}). We can see clearly in this figure the three different behaviors for the mass of the black hole discussed above: if $M < M_{CMB}$, the mass will decrease to zero (until complete evaporation), if $M > M_{CMB}$, the mass will diverge, and if $M = M_{CMB}$, the mass will remain constant. The $M = 1$ in the figure plays the role of $M_{CMB}$, and separates the other two kinds of solutions. The total time that takes for the mass to diverge or reach zero depends of the initial mass $M = M_0$.

To obtain particular solutions to the differential equation (\ref{eq:dJm}), we will analyze its asymptotic behavior, because now its not so simple as equation (\ref{eq:Mt}). The energy emission is associated to the first term $- \frac{a}{M^2}$, while the absorption is related to the second term $b M^2$. So, we have the following asymptotic solutions:
\begin{align}
  1st\hspace{0.1cm}\textnormal{term:} &\hspace{0.1cm} \frac{d M}{d t} = - \frac{a}{M^2} \nonumber\\
  & M (t) = M_0 \left(1 - \frac{t}{t_V}\right)^{1/3} , \hspace{0.1cm} t \in [0, t_V) \label{eq:M=vazio}\\
  2nd\hspace{0.1cm}\textnormal{term:} &\hspace{0.1cm} \frac{d M}{d t} = b M^2 \nonumber\\
  & M (t) = \frac{M_0}{\left( 1 - b M_0 t \right)}, \hspace{0.1cm} t \in \left[0, \frac{1}{b M_0}\right) \label{eq:Msegundotermob}
\end{align}
Notice that equation (\ref{eq:M=vazio}) is identical to solution (\ref{eq:M(t)}) for the empty space. Taking the \emph{numerical solution} of differential equation \eqref{eq:dJm} and comparing with the asymptotic solutions (\ref{eq:M=vazio}) and (\ref{eq:Msegundotermob}), we can analyze how the masses of black holes change with time.

\begin{figure}[h!]
 \centering
 \includegraphics[scale=0.34]{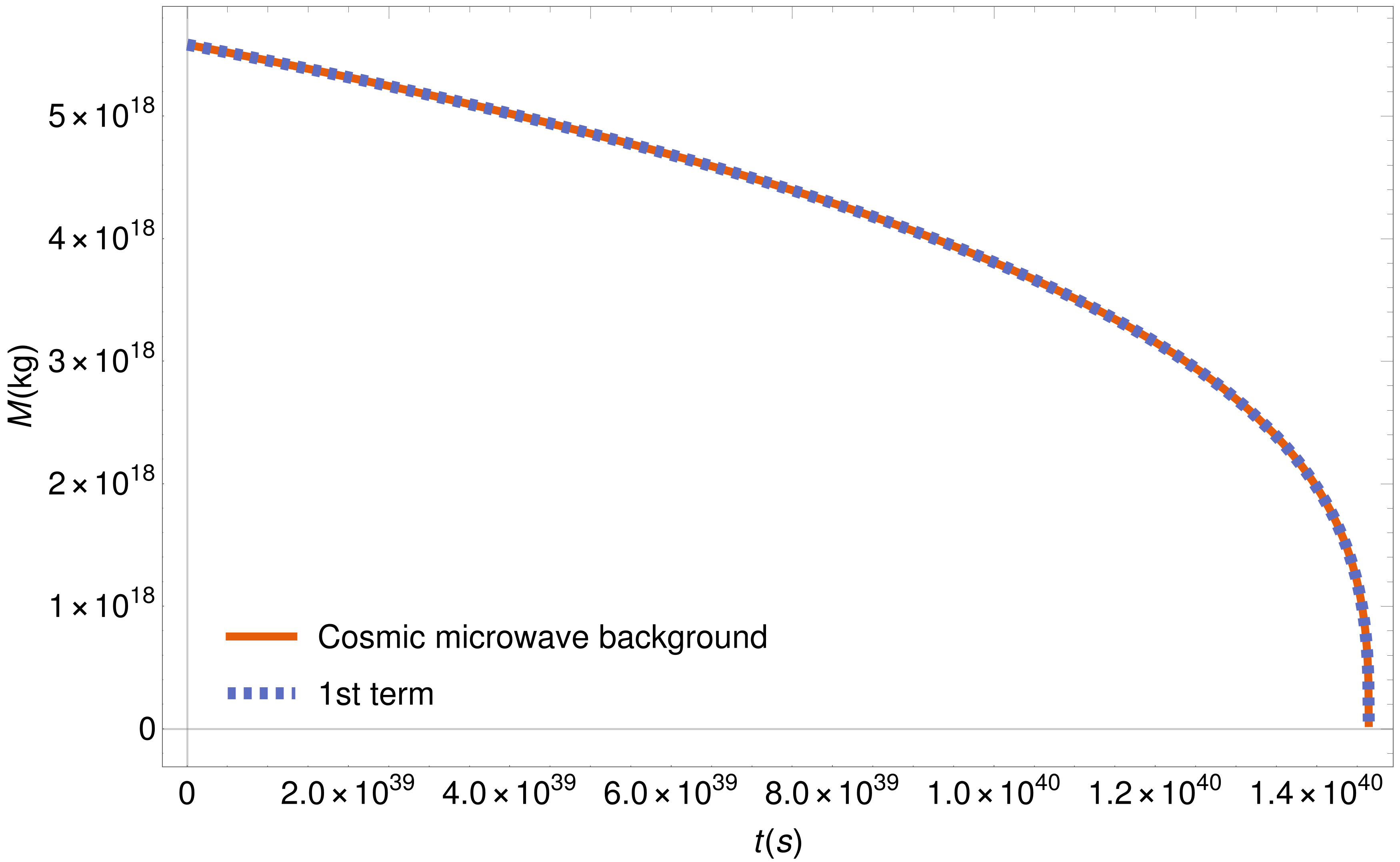}
 \caption{\label{fig:EX1Tcte} Analytical asymptotic solution (for the 1st term) and complete numerical solution for a Schwarzschild black hole with $M = M_0^{hip} \simeq 5.6 \cdot 10^{18} kg$ of Hyperion, immersed in cosmic microwave background.}
\end{figure}

In the figure \ref{fig:EX1Tcte} we used as initial condition Hyperion mass $M = M_0^{hip} \simeq 5.6 \cdot 10^{18} kg$, smaller than the critical mass $M_{CMB}$ \eqref{eq:Mrcf}. Thus, comparing the \emph{complete numerical solution} (of equation \eqref{eq:dJm}) with \emph{1st term analytical solution} (\ref{eq:M=vazio}), we verify the superposition of these two solutions and the decreasing of mass of this black hole with $M_0^{hip}$.
Also, notice that the time required for this black hole to evaporate is exactly the lifetime for Hyperion, calculated in table \ref{tab:tv}, as expected by \emph{1st term analytical solution} (\ref{eq:M=vazio}).

\begin{figure}[h!]
 \centering
 \includegraphics[scale=0.34]{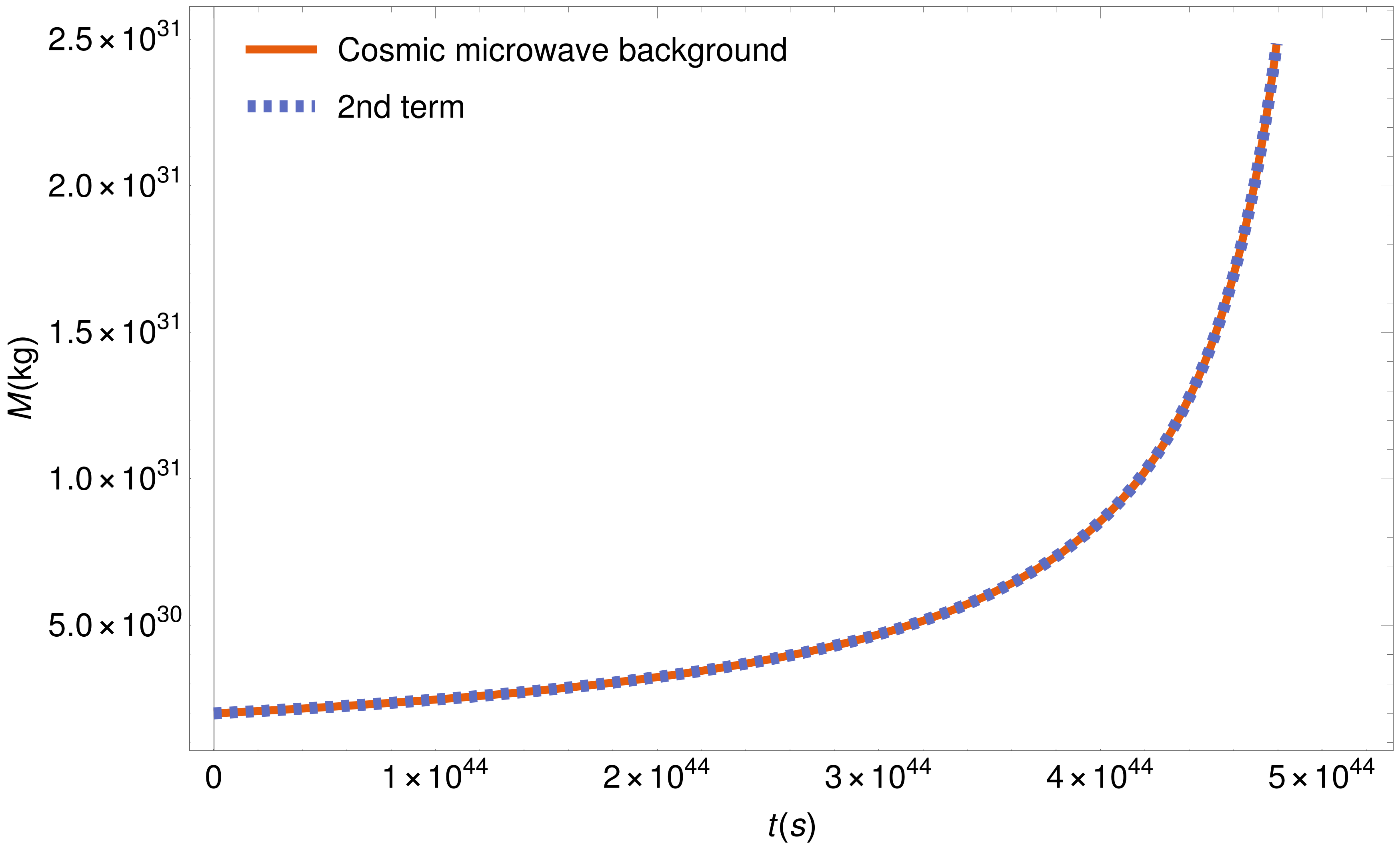}
 \caption{\label{fig:EX2Tcte} Analytical asymptotic solution (for the 2nd term) and complete numerical solution for a Schwarzschild black hole with $M = M_0^{\bigodot} \simeq 2.0 \cdot 10^{30} kg$ of Sun, immersed in cosmic microwave background.}
\end{figure}

In Figure \ref{fig:EX2Tcte} we considered as initial condition a Schwar-zschild black hole with Sun mass $M = M_0^{\bigodot} \simeq 2.0 \cdot 10^{30} kg$, greater than the critical mass $M_{CMB}$ \eqref{eq:Mrcf}. Again, comparing the \emph{complete numerical solution}, of equation \eqref{eq:dJm}, with the \emph{2nd term analytical solution} (\ref{eq:Msegundotermob}), we can see the perfect accordance between this asymptotic solutions and the numerical solution. Moreover, the maximum time achieved by both curves in this figure (i.e., the time required for the mass to diverge) is exactly the upper limit for the time interval in equation (\ref{eq:Msegundotermob}), that for the Sun is $t_C^{\bigodot} = \frac{1}{b M^{\bigodot}_0} \simeq 5.21 \cdot 10^{44} s$.

Therefore, according to the initial mass, these asymptotic solutions (\ref{eq:M=vazio}) and (\ref{eq:Msegundotermob}) model very well the behavior of the masses of Schwarzschild black holes. The equation (\ref{eq:M=vazio}) can be used when $M < M_{CMB}$, and (\ref{eq:Msegundotermob}) when $M > M_{CMB}$, because in both cases one of the two terms in the RHS of equation (\ref{eq:dJm}) will be dominant over the other one, and this approach will be better the further away the initial mass $M = M_0$ is from $M_{CMB}$ (the convergence will be faster).

Finally, notice that the intervals considered for the time in equations (\ref{eq:M=vazio}) and (\ref{eq:Msegundotermob}) are open in $t_V$ and $\frac{1}{bM_0}$, and the reason is that the mass goes to zero or infinity, respectively. Actually, there is no consensus about what happens in these two limits, but we can safely assume that the approximation used in this paper, i.e., perfect black body emitting radiation accordingly to Stefan-Boltzmann law, will no longer be valid, and other physics phenomena must be included in the discussion, but that is beyond the scope of this paper. More details about this discussion can be found in the section \ref{sec:disc}.

%------------------------------------------------------------------------

\section{Black hole and the thermal evolution of the universe}
\label{sec:ETU}

We know that the universe evolve in time and, thus, there is a thermal evolution of the matter and the energy content, mainly for long time scales. In this way, this thermal evolution is an important point to be considered for the mass evolution of a Schwarzschild black hole. The details of the calculations and approximations used in the present section are found in appendix \ref{sec:etu}.

In this analysis we will consider that the black hole exchange energy with the cosmic microwave background and that the CMB temperature evolves in time according to different universe eras: \emph{radiation}, \emph{matter} and \emph{dark energy} \footnote{The time evolution of the universe is marked by three eras, i.e., three time intervals that are named according to the dominant content. Considering the ``big bang'' as time's origins, the eras and its time interval are: \emph{radiation era} $t \in (0, 9.3 \cdot 10^{11}) s$, \emph{matter era} $t \in (9.3 \cdot 10^{11}, 3.7 \cdot 10^{17}) s$ and \emph{dark energy era} $t \in (3.7 \cdot 10^{17}, \infty) s$. For more information see appendix \ref{sec:etu}.}. Hence, the thermal evolution of the CMB, in accordance with (\ref{eq:Traderas}), is
\begin{align}
  T_R (t) & = \frac{T_{CMB}}{\left(\frac{32 \pi G}{3} \rho_{R_0}\right)^{1/4}} \frac{1}{t^{1/2}}, \hspace{0.1cm} t \in (0, t_{p_1}) \label{eq:TR}\\
  T_M (t) & = \frac{T_{CMB}}{\left(6 \pi G \rho_{M_0}\right)^{1/3}} \frac{1}{t^{2/3}} , \hspace{0.1cm} t \in (t_{p_1}, t_{p_2}) \label{eq:TM}\\
  T_{\Lambda} (t) & = T_{CMB} \textnormal{Exp}\hspace{-0.05cm} \left[\hspace{-0.05cm}- \hspace{-0.08cm}\left(\hspace{-0.1cm}\frac{8 \pi G}{3} \rho_{\Lambda_0}\hspace{-0.1cm}\right)^{\hspace{-0.1cm}1/2} \hspace{-0.2cm}t\hspace{-0.05cm}\right] , t \in (t_{p_2}, \infty) \label{eq:TE}
\end{align}
where $t_{p_1} = \left[\frac{3}{32 \pi G \rho_{R_0}} \left(\frac{\rho_{R_0}}{\rho_{M_0}}\right)^4\right]^{1/2} \simeq 9.3 \cdot 10^{11} s$ is the limit time for the \emph{radiation era}, $t_{p_2} = \frac{1}{(6 \pi G \rho_{\Lambda_0})^{1/2}} \simeq 3.7 \cdot 10^{17} s$ is the limit time for the \emph{matter era}, $T_{CMB}$ represents temperature of cosmic microwave background today and $\rho_{R_0}$, $\rho_{M_0}$ and $\rho_{\Lambda_0}$ represents radiation, matter and dark energy densities today, respectively.

The procedure here will be the same used previously in the section \ref{sec:rad}, i.e., we are going to use the Stefan-Boltzmann law as a thermal current between the Schwarzschild black hole and the cosmic microwave background in the way that
\begin{equation}
  J_i = \sigma A \left( T^4 - T^4_i \right) ,
\end{equation}
with $i =$ $R$ (radiation), $M$ (matter) or $\Lambda$ (dark energy). Again, the black hole will emit ($T > T_{i}$) or absorb ($T < T_{i}$) energy, and the CMB will behave as a thermal reservoir\footnote{Even thought the CMB temperature is changing with time, we are considering that this change does not interfere with the energetic exchange between the CMB and the black hole, i. e., the CMB is still a thermal reservoir.}. Thus, substituting the temperatures (\ref{eq:TR}-\ref{eq:TE}) and already transforming $J = - \frac{d (M c^2)}{d t}$, we have
\begin{align}
 \textnormal{Radiation:}\hspace{0.3cm}\frac{d M}{d t} & = - \frac{a}{M^2} + b\hspace{0.1cm}a_R \frac{M^2}{t^2} \label{eq:dMR}\\
 \textnormal{Mater:}\hspace{0.3cm}\frac{d M}{d t} & = - \frac{a}{M^2} + b\hspace{0.1cm}a_M \frac{M^2}{t^{8/3}} \label{eq:dMM}\\
 \textnormal{Dark energy:}\hspace{0.3cm} \frac{d M}{d t} & = - \frac{a}{M^2} + b\hspace{0.1cm}M^2 e^{- a_{\Lambda} t} \label{eq:dMD} 
\end{align}
with $a = \frac{\hbar c^4}{15360 \pi G^2} \simeq 4.0 \cdot 10^{15} \frac{kg^3}{s}$, $b = \frac{4 \pi^3 k_B^4 G^2 T^4_{CMB}}{15 \hbar^3 c^8} \simeq 9.6 \cdot 10^{- 76} \frac{1}{s \cdot kg}$, $a_R = \frac{1}{\left(\frac{32 \pi G}{3}\rho_{R_0}\right)} \simeq 9.6 \cdot 10^{38} s^2$, $a_M = \frac{1}{\left(6 \pi G \rho_{M_0}\right)^{4/3}} \simeq 2.0 \cdot 10^{47} s^{8/3}$ and $a_{\Lambda} = \left(\frac{128 \pi G}{3} \rho_{\Lambda_0}\right)^{1/2} \simeq 7.2 \cdot 10^{- 18} s^{- 1}$. The equations (\ref{eq:dMR}-\ref{eq:dMD}) are more complicated than equation (\ref{eq:dJm}) because we have direct dependence in $t$. In this way, there is no analytical solution.

\begin{figure}[h!]
 \centering
 \includegraphics[scale=0.37]{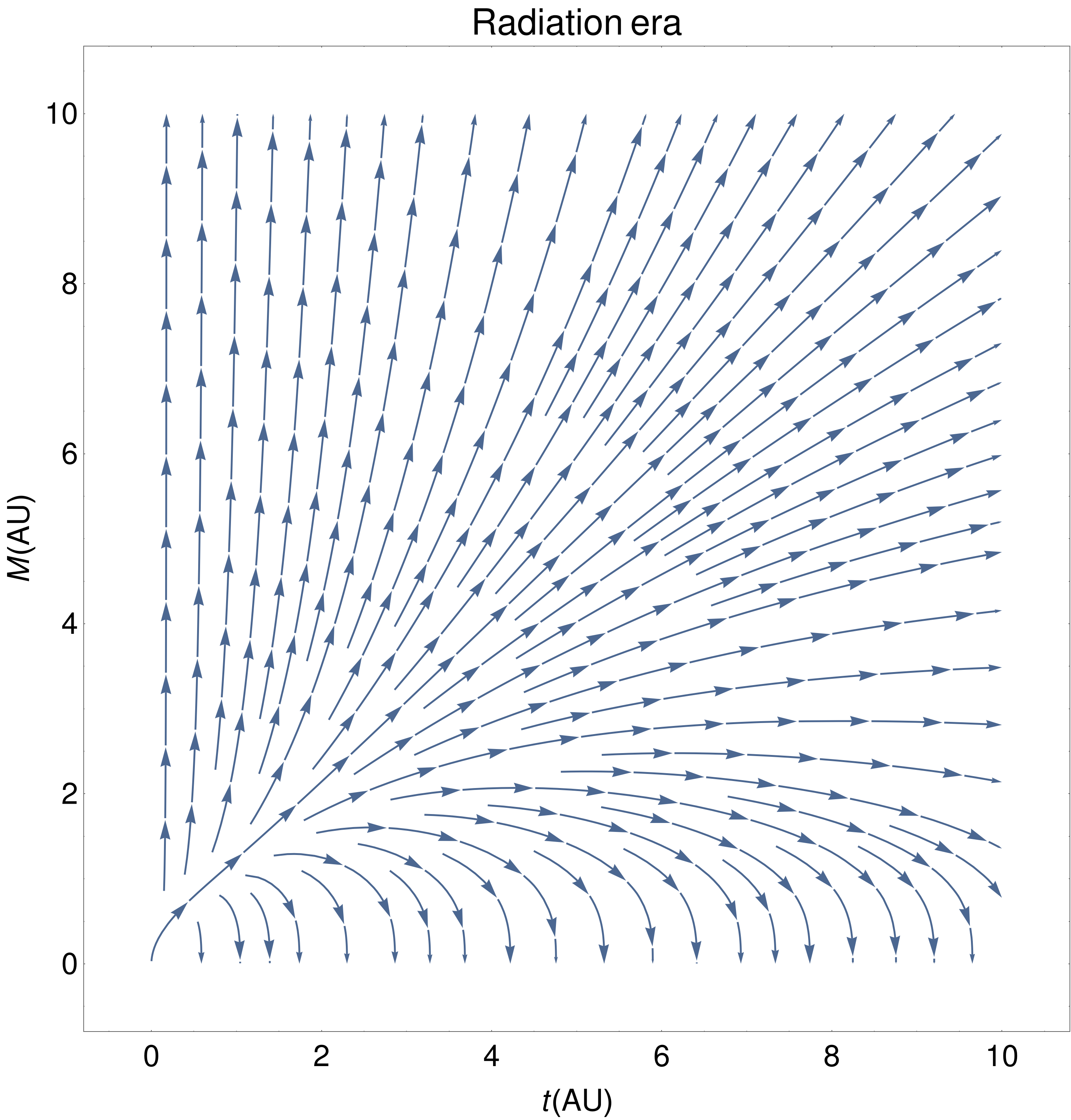}
 \caption{\label{fig:streamplotsRMD} \emph{Stream plot} (mass {\em versus} time) for a Schwarzschild black hole exchanging energy with CBM in the radiation era. Notice that we used arbitrary units (AU).}
\end{figure}

In the Figure \ref{fig:streamplotsRMD} below we have constructed the {\em stream plot} for the differential equation (\ref{eq:dMR}), in arbitrary units. The goal of this figure is to show the general behavior, not only for solution of the equation (\ref{eq:dMR}), but for all solutions of equations (\ref{eq:dMR}-\ref{eq:dMD}). The reason is that these three differential equations have the same general behavior for the mass evolution of Schwarzschild black holes. Note that now these equations do not have a critical mass, as the $M_{CMB}$ \eqref{eq:Mrcf} in the section \ref{sec:rad}. Hence, we can see two different behaviors: the mass will decrease to zero, until complete evaporation (even though it can increase in the beginning) or the mass will diverge, accordingly to its initial mass and universe era. Finally, there is no constant mass behavior because we have the time evolution in all the three cases.

As we have made in section \ref{sec:rad}, we can analyze the asymptotic behaviors for equations (\ref{eq:dMR}-\ref{eq:dMD}). These three equations have two terms that leads to two different asymptotic solutions. The \emph{1st term} of differential equations (\ref{eq:dMR}-\ref{eq:dMD}) (equal for the three equations) can be written as 
\begin{align}
  & 1^{srt} \hspace{0.1cm}\textnormal{term:} \hspace{0.5cm}\frac{d M_i }{dt} = - \frac{a}{M^2_i} \nonumber\\
  & \hspace{0.3cm}\Rightarrow\hspace{0.3cm} M_i (t) = M_{0_i} \left[ 1 - \frac{\left(t - t_{0_i}\right)}{t_{V_i}}\right]^{1/3} , \label{eq:1TERMOERAS}
\end{align}
where the index $i = R, M$ or $\Lambda$ represents respectively radiation, matter or dark energy eras, $M_i (t_{0_i}) = M_{0_i}$ is the initial mass for the black hole in each era, $t_{0_i}$ represents the initial time in each era ($t_{0_R} = 0$\footnote{Observe that, physically, $t_{0_i} = 0$ for radiation era. However, as we have solved the expressions numerically, we have fixed a initial time, being small but not null. The choice was $t_{0_R} \equiv 10^{-3} s$. \label{fn:10-3rad}} for radiation era, $t_{0_M} = t_{p_1} \simeq 9.3 \cdot 10^{11} s$ for matter era and $t_{0_{\Lambda}} = t_{p_2} \simeq 3.7 \cdot 10^{17} s$ for dark energy era), and $t_{V_i} = 5120 \pi \frac{G^2}{\hbar c^4} M_{0_i}^3$ represents the lifetime of black hole\footnote{Note that the lifetime $t_{V_i}$ just refers to the initial mass of the black hole, i.e., it is independent of any parameter relative to the universe eras. This happens because the equations (\ref{eq:dMR}-\ref{eq:dMD}) have the same 1st term.}. The equation \eqref{eq:1TERMOERAS} has the intervals:
\begin{align}
  t & \in ( t_{0_i}, t_{0_i} + t_{V_i} ) , \hspace{0.3cm} \textnormal{if} \hspace{0.3cm} t_{0_i} + t_{V_i} < t_{\textnormal{end of era}}^i \label{eq:evapintervaloera1}\\
  t & \in ( t_{0_i}, t^i_{\textnormal{end of era}} ) , \hspace{0.3cm} \textnormal{if} \hspace{0.3cm} t_{0_i} + t_{V_i} > t_{\textnormal{end of era}}^i \label{eq:evapintervaloera}
\end{align}
where $t_{\textnormal{end of era}}^i$ is the final time of each era ($t_{\textnormal{end of era}}^R = t_{p_1} \simeq 9.3 \cdot 10^{11} s$ for the radiation era, $t_{\textnormal{end of era}}^M = t_{p_2} \simeq 3.7 \cdot 10^{17} s$ for the matter era and $t_{\textnormal{end of era}}^{\Lambda} \sim \infty \hspace{0.05cm}s$ for the dark energy era). Moreover, the equation (\ref{eq:evapintervaloera}) states that if the black holes lifetime $t_{V_i}$ exceeds the limit of the considered era, we need to consider the solution relative to the next era, using its final mass (relative to the previous era) as initial mass for the next era. Besides that, the \emph{1st term} is still related to \emph{energy emission} for black holes.

The \emph{2nd term} of the differential equations (\ref{eq:dMR}-\ref{eq:dMD}) are different and they can be written as
\begin{align}
  &2nd \hspace{0.1cm}\textnormal{Radiation:} \hspace{0.3cm}\frac{d M}{dt} = b\hspace{0.05cm}a_R \frac{M^2}{t^2} ,\nonumber\\
  & \hspace{0.3cm}\Rightarrow\hspace{0.3cm} M(t) = \frac{M_{0_R}\hspace{0.05cm}t_{0_R}\hspace{0.05cm}t}{t_{0_R}\hspace{0.05cm}t + M_{0_R}\hspace{0.05cm}b\hspace{0.05cm}a_R \left( t_{0_R} - t \right)} , \label{eq:2rad}\\
  &2nd \hspace{0.1cm}\textnormal{Matter:} \hspace{0.3cm}\frac{d M}{dt} = b\hspace{0.05cm}a_M \frac{M^2}{t^{8/3}} ,\nonumber\\
  & \hspace{0.3cm}\Rightarrow\hspace{0.3cm} M(t) = \frac{5\hspace{0.05cm}M_{0_M}\hspace{0.05cm}t_{0_M}^{5/3}\hspace{0.05cm}t^{5/3}}{5\hspace{0.05cm}t_{0_M}^{5/3}\hspace{0.05cm}t^{5/3} + 3\hspace{0.05cm}b\hspace{0.05cm}a_M\hspace{0.05cm}M_{0_M} \left(t_{0_M}^{5/3} - t^{5/3}\right)} , \label{eq:2mat}\\
  &2nd \hspace{0.1cm}\textnormal{Dark energy:} \hspace{0.3cm}\frac{d M}{dt} = b\hspace{0.05cm}M^2 e^{- a_{\Lambda} t} ,\nonumber\\
  & \hspace{0.3cm}\Rightarrow\hspace{0.3cm} M(t) = \frac{M_{0_{\Lambda}} a_{\Lambda}}{a_{\Lambda} + M_{0_{\Lambda}} b \left( e^{- a_{\Lambda} t} - e^{- a_{\Lambda} t_{0_{\Lambda}}} \right)} , \label{eq:2dark}
\end{align}
where $M_{0_R}$, $M_{0_M}$, $M_{0_{\Lambda}}$, $t_{0_R}$, $t_{0_M}$ and $t_{0_{\Lambda}}$ have the same values interpreted and discussed after equation (\ref{eq:1TERMOERAS}). The respective intervals are
\begin{align}
  & \textnormal{Radiation:} \hspace{0.3cm}t \in \left( t_{0_R}, t_{C}^{R} \right) ,\nonumber\\
  & t_C^{R} = \frac{b\hspace{0.05cm}a_R\hspace{0.05cm}M_{0_R}\hspace{0.05cm}t_{0_R}}{\left(b\hspace{0.05cm}a_R\hspace{0.05cm}M_{0_R} - t_{0_R} \right)} \hspace{0.2cm}\textnormal{if}\hspace{0.2cm} t_C^{R} < t^R_{\textnormal{end of era}} , \label{eq:int2rad}\\
  & \textnormal{Matter:} \hspace{0.3cm}t \in \left(t_{0_M}, t_{C}^{M} \right) ,\nonumber\\
  & t_C^{M} = \frac{\left(3\hspace{0.05cm}b\hspace{0.05cm}a_M\hspace{0.05cm}M_{0_M}\right)^{3/5} t_{0_M}}{\left(3\hspace{0.05cm}b\hspace{0.05cm}a_M\hspace{0.05cm}M_{0_M} - 5\hspace{0.05cm}t_{0_M}^{5/3}\right)^{3/5}} \hspace{0.2cm}\textnormal{if}\hspace{0.2cm} t_C^{M} < t^M_{\textnormal{end of era}} , \label{eq:int2mat}\\
  &\textnormal{Dark energy:} \hspace{0.25cm}t \in \left(t_{0_{\Lambda}}, t_{C}^{\Lambda} \right) ,\nonumber\\
  & t_{C}^{\Lambda} = - \frac{1}{a_{\Lambda}} \ln \left( e^{- a_{\Lambda} t_{0_{\Lambda}}} - \frac{a_{\Lambda}}{b M_{0_{\Lambda}}}\right) \hspace{0.15cm}\textnormal{if}\hspace{0.15cm} t_C^{\Lambda} < t^{\Lambda}_{\textnormal{end of era}} , \label{eq:int2dark}
\end{align}
where $t_C^i$, for $i = R, M$ or $\Lambda$, represents the critical time (inside each era) for which the mass of the black hole diverges and $t_{\textnormal{end of era}}^i$ has the same interpretation and the same values according to discussion after equations (\ref{eq:evapintervaloera1}) and (\ref{eq:evapintervaloera}). Thus, we have a second case for the time interval of those three eras given by
\begin{equation}
 t \in ( t_{0_i}, t^i_{\textnormal{end of era}} ) , \hspace{0.3cm} \textnormal{if} \hspace{0.3cm} t_{C}^i > t_{\textnormal{end of era}}^i . \label{eq:crescinterasparou}
\end{equation}
The equation \eqref{eq:crescinterasparou} state that, if the critical time $t_C^i$ exceeds the limit of the considered era, we need to consider the solution relative to the next era, using its final mass (relative to the previous era) as initial mass for the next era. Moreover, the \emph{2nd term} is related to \emph{energy absorption} for black holes.
Again, we are going to compare the asymptotic solutions to the complete numerical solutions.

\begin{figure}[h!]
  \centering
  \includegraphics[scale=0.277]{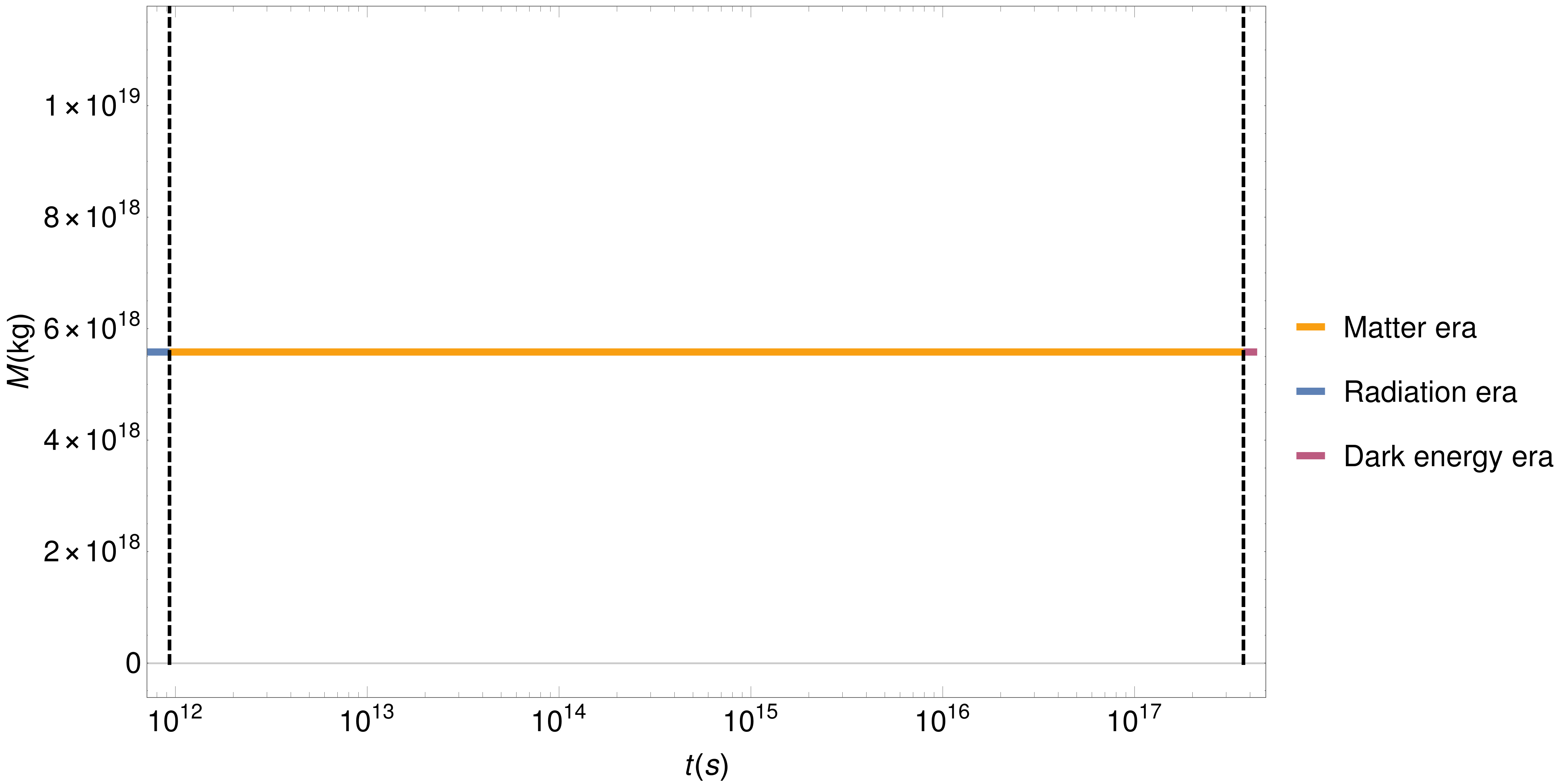}\\
  \caption{\label{fig:hiperas} Complete numerical solution to time evolution of a black hole with initial mass $M^{Hip}_0 \simeq 5.6 \cdot 10^{18} kg$ of Hyperion for radiation, matter and dark energy eras. Note that we have used the logarithm scale in the horizontal axis.}
\end{figure}

\begin{figure}[h!]
  \centering
  \includegraphics[scale=0.277]{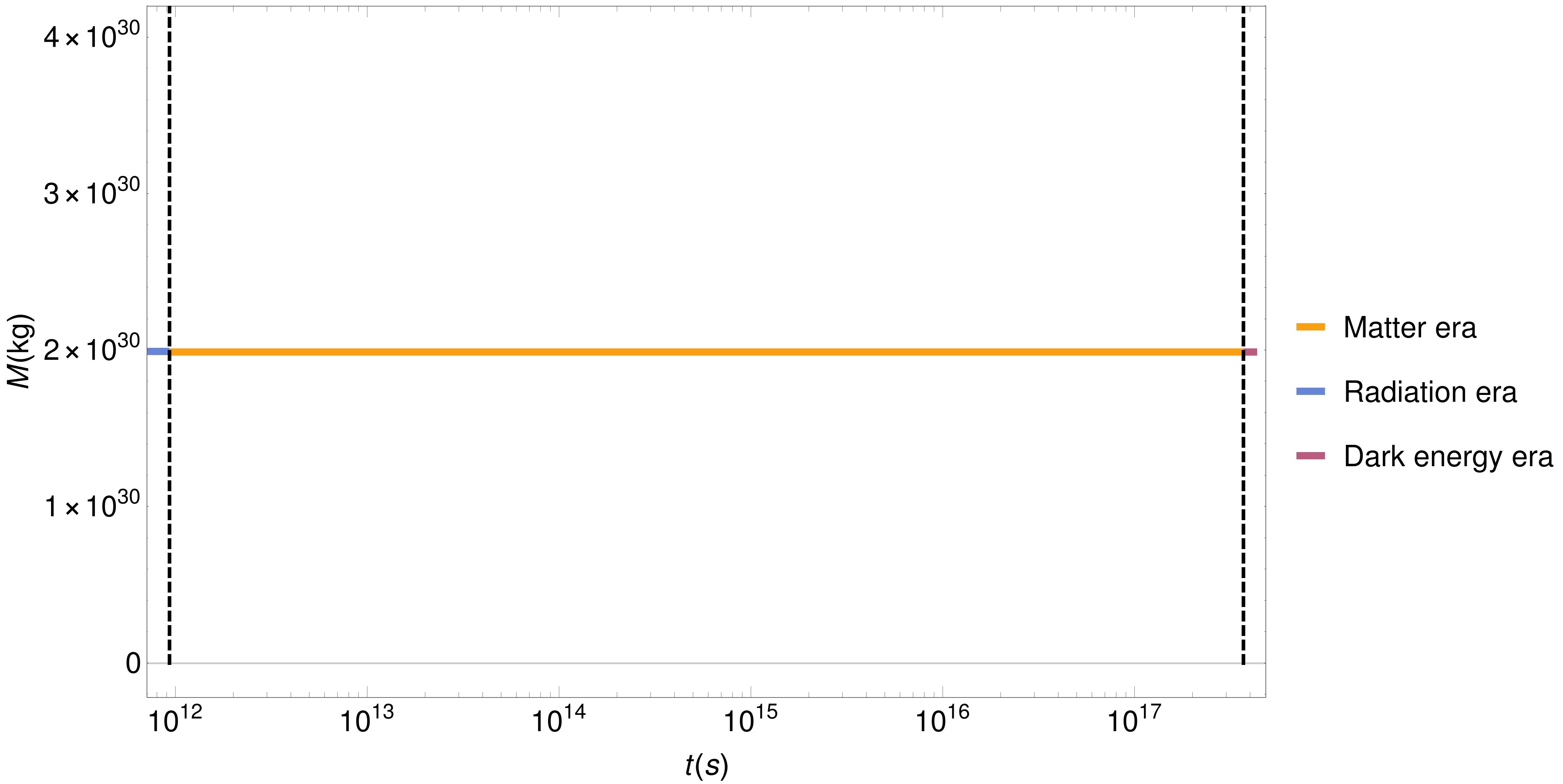}\\
  \caption{\label{fig:soleras} Complete numerical solution to time evolution of a black hole with initial mass $M_0^{\bigodot} \simeq 2.0 \cdot 10^{30} kg$ of Sun for radiation, matter and dark energy eras. Note that we have used the logarithm scale in the horizontal axis.}
\end{figure}

We can analyze the mass of the black hole using the \emph{analytical asymptotic solution} and the \emph{numerical solution} of equations (\ref{eq:dMR}-\ref{eq:dMD}). First, we are going to study the behavior of black holes with initial masses of Hyperion and Sun. As we can see in the Figures \ref{fig:hiperas} and \ref{fig:soleras}, we verify that, for these two cases, the mass remains approximately constant within each universe era. The vertical dotted lines represent the intersection between two respective eras: $t_{R \& M} = t_{p_1} \simeq 9.3 \cdot 10^{11} s$ between radiation and matter eras and $t_{M \& \Lambda} = t_{p_2} \simeq 3.7 \cdot 10^{17} s$ between matter and dark energy eras. Note that we have used the logarithm scale in the horizontal axis to get a better representation for the three eras (the time scales of radiation and dark energy eras are small when compared to the matter era scale).

Thus, for those initial conditions we expect the decrease of the black hole mass for very long periods of time. It is interesting to note that, comparing the two cases with initial Sun mass ($M_0 = M_{\bigodot}$), for constant CMB temperature and CMB temperature varying accordingly to the universe eras, in the former case the mass will increase and, in the latter case it will decrease.

\begin{figure}[h!]
  \centering
  \includegraphics[scale=0.34]{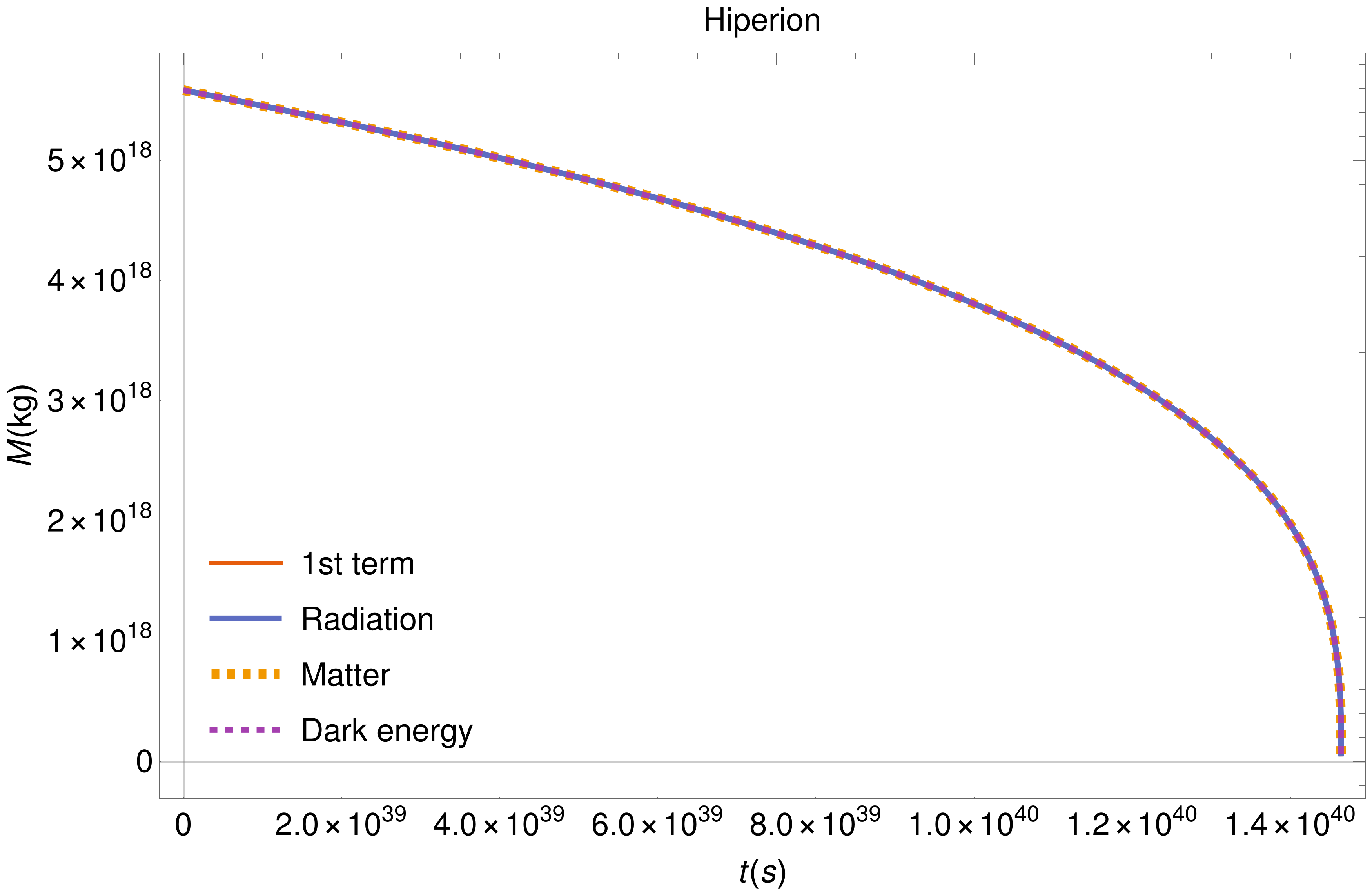}\vspace{0.2cm}
  \includegraphics[scale=0.34]{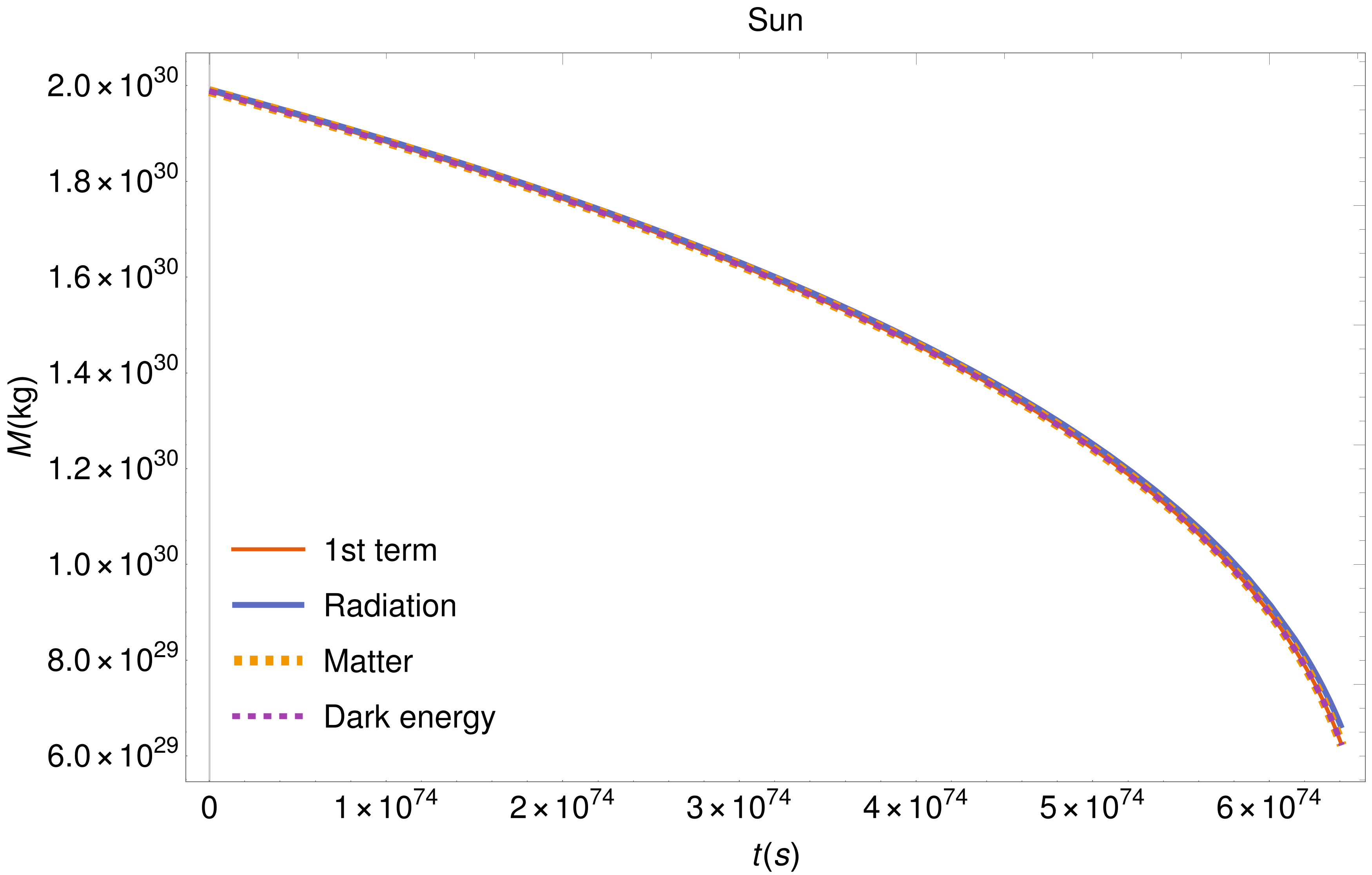}
  \caption{\label{fig:hipsoldectotal} Mass evolution comparison of black hole immersed in cosmic microwave radiation with temperature evolving according to the universe eras with initial masses $M_0^{hip} \simeq 5.6 \cdot 10^{18} kg$ of Hyperion and $M_0^{\bigodot} = 2.0 \cdot 10^{30} kg$ of Sun. The plots show the analytical solution of the \emph{1st term} and \emph{complete numerical solution} to each era.}
\end{figure}

Now we can analyze the behavior of masses of Hyperion and Sun, still using \emph{analytical asymptotic solution} and {\em complete numerical solution}, but extrapolating the time intervals of each era, i.e., taking them as $t \in (0, t_V)$, where $t_V$ is the lifetime of each body ($t_v^{Hip} \simeq 1.5 \cdot 10^{40} s$ e $t_v^{\bigodot} \simeq 6.6 \cdot 10^{74} s$). In this way we can compare those solutions to asymptotic analytical solution of \emph{1st term} (\ref{eq:1TERMOERAS}) of differential equations (\ref{eq:dMR}-\ref{eq:dMD}), which is equivalent to the solution of \emph{empty space} \eqref{eq:M(t)}. This comparison is made in the Figure \ref{fig:hipsoldectotal}, where we see the superposition between the {\em complete numerical solution} to each era and the \emph{1st term} analytical solution. We can see that the mass are decreasing and consequently the temperature increasing, leading to its evaporation due to radiation emission. Hence, the asymptotic solution of \emph{1st term} provides perfectly the behavior of the mass of a black hole for these specific initial conditions.

\begin{table}[h!]
\begin{center}
\caption{\label{tab:mtv}Black hole masses with lifetime equal to the era duration: radiation, matter and dark energy.}
\begin{tabular}{c|cc}
  \cline{2-3}
   & $\mathbf{t_V}$ \textbf{(s)} & \textbf{M (kg)} \\
   \cline{1-3}
   \multicolumn{1}{c|}{\textbf{Radiation era}}  & $9.3 \cdot 10^{11}$ & $2.2 \cdot 10^9$ \\
   \cline{1-3}
   \multicolumn{1}{c|}{\textbf{Matter era}} & $3.7 \cdot 10^{17}$ & $1.6 \cdot 10^{11}$ \\
   \cline{1-3}
   \multicolumn{1}{c|}{\textbf{Dark energy era}} & $6.6 \cdot 10^{16}$ & $9.2 \cdot 10^{10}$ \\
   \cline{1-3}
\end{tabular}
\end{center}
\end{table}

Thus, for those initial conditions we expect the decrease of the black hole mass for very long periods of time (in the dark energy era). It is interesting to note that, comparing the two cases with initial Sun mass ($M_0 = M_{\bigodot}$), for constant CMB temperature and CMB temperature varying accordingly to the universe eras, in the former case the mass will increase until diverge and in the latter case it will decrease until complete evaporation.

\begin{figure}[h!]
  \centering
  \includegraphics[scale=0.34]{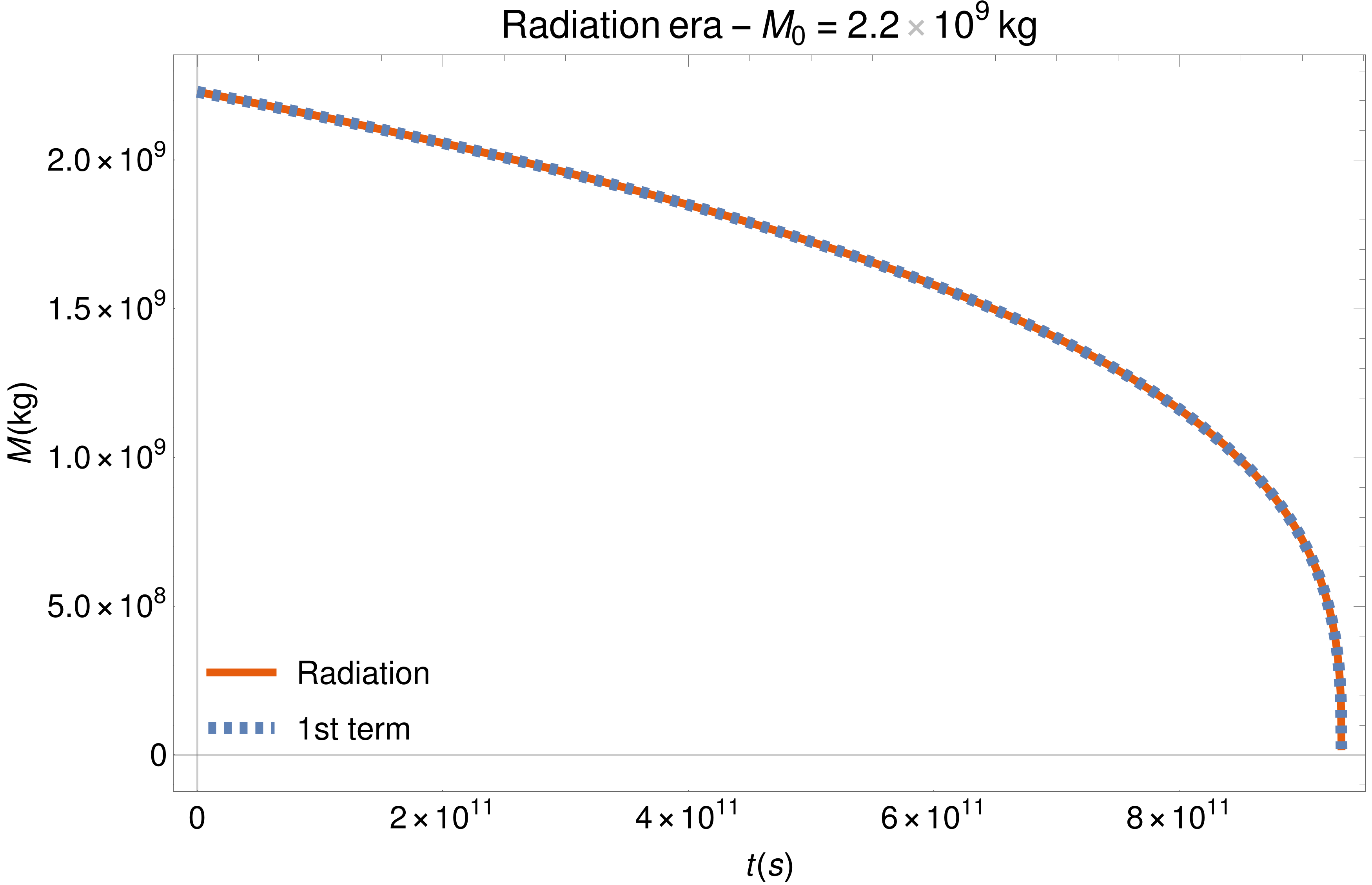}\\
  \includegraphics[scale=0.34]{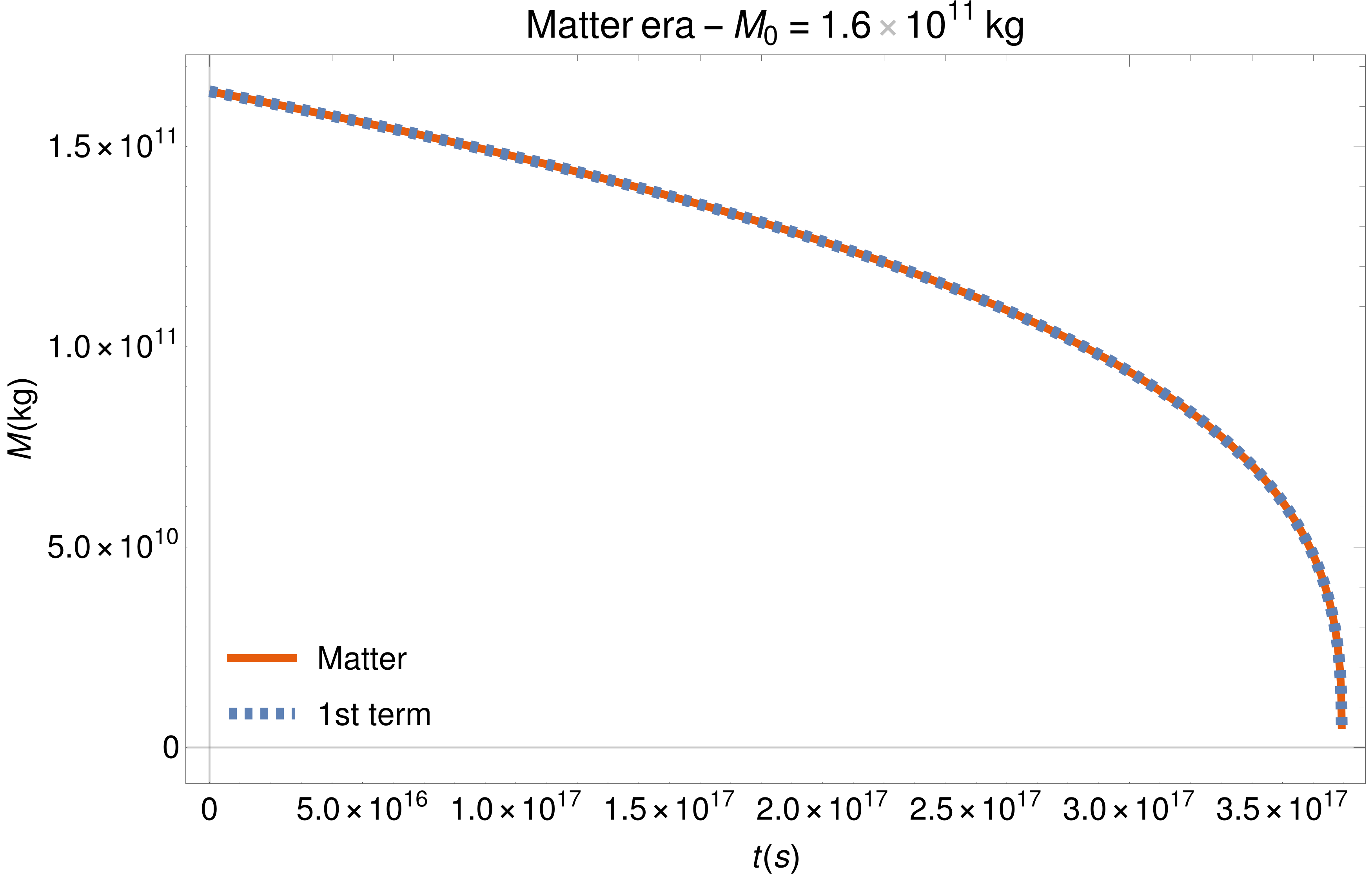}\\
  \hspace{0.4cm}\includegraphics[scale=0.34]{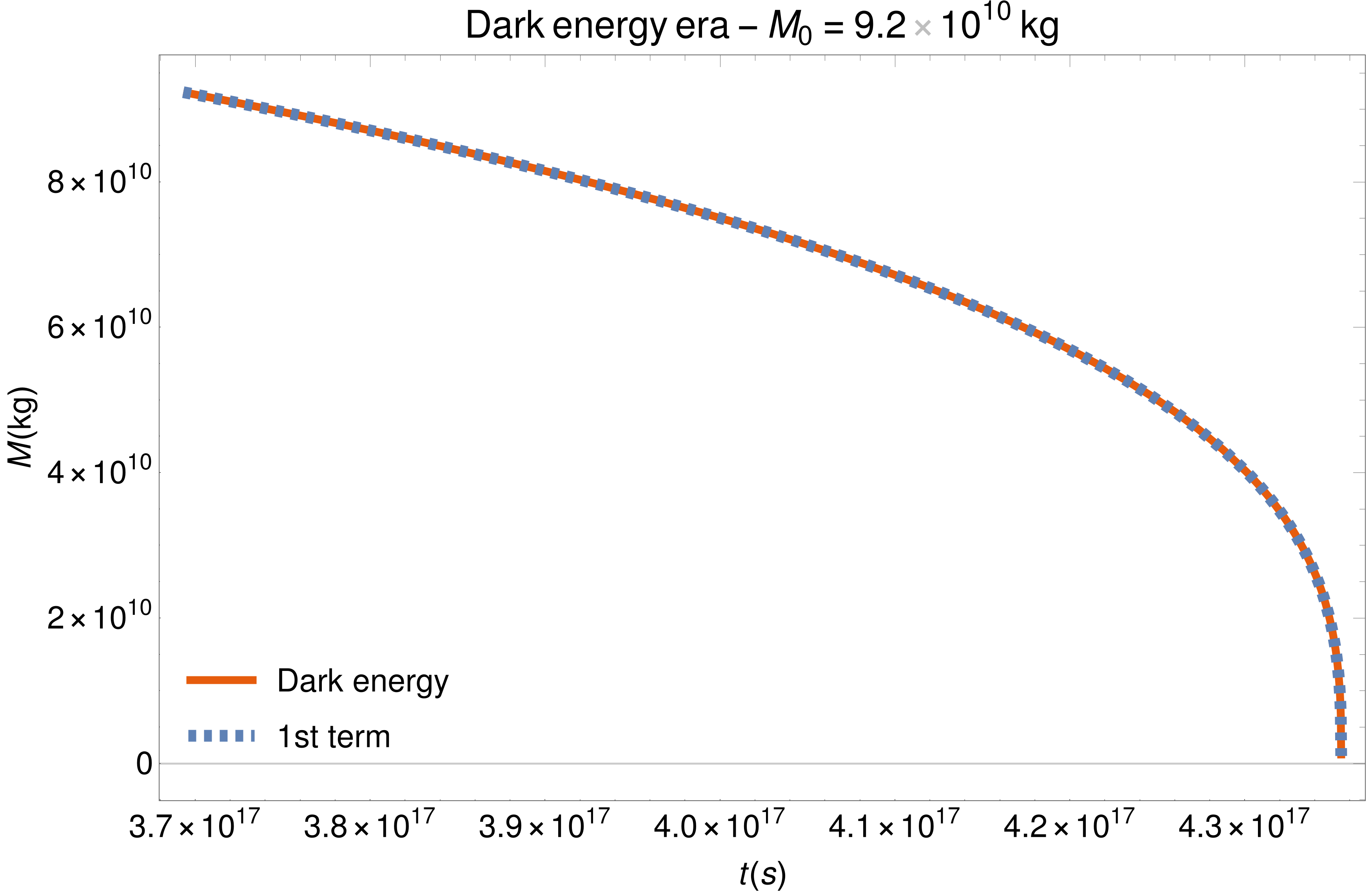}
  \caption{\label{fig:RMEevapc}Mass evolution comparison of a black hole immersed in cosmic microwave background radiation with temperature evolving according to the universe eras with initial masses equal to the masses of black holes with lifetime according to the time interval of each universe era (given by Table \ref{tab:mtv}). The plots show the analytical solution of the {\em 1st term} and the \emph{complete numerical solution} to each era, respectively.}
\end{figure}

Following the idea of complete evaporation of a black hole, we computed in Table \ref{tab:mtv} the initial masses of black holes with lifetime equal to the duration of the universe eras, and we considered, for convenience, the end of dark energy era as the universe lifetime (today date). Then we verified the expected behavior, of evaporation, for those masses using \emph{complete numerical solution} of equations (\ref{eq:dMR}-\ref{eq:dMD}) comparing with analytical solution of \emph{1st term} (\ref{eq:1TERMOERAS}). As we can see in the Figure \ref{fig:RMEevapc}, these plots always  match, confirming the previous presented idea. Therefore, for black holes with masses smaller than the masses founded here we always have the complete evaporation, in each era respectively.

We also have found the initial mass, for each era, that makes the black hole to increase its mass initially, i.e., the mass of what we expect to follow the ``unstable equilibrium'' between the two situations: emission and absorption of radiation. We have estimated these values imposing that the second derivative in time of equations (\ref{eq:dMR}-\ref{eq:dMD}) is null, and these values can be seen in the Table \ref{tab:mmincres}. See that, one more time, we have used the initial time $t_0$ for the radiation era as explained in the footnote \ref{fn:10-3rad}.

\begin{table}[h!]
\begin{center}
\caption{\label{tab:mmincres}Minimum black hole masses that present initially increasing behavior for each universe eras.}
\begin{tabular}{c|cc}
  \cline{2-3}
   & $\mathbf{t_0}$ \textbf{(s)} & \textbf{M (kg)} \\
   \cline{1-3}
   \multicolumn{1}{c|}{\textbf{Radiation era}}  & $10^{-3}$ & $1.1 \cdot 10^{33}$ \\
   \cline{1-3}
   \multicolumn{1}{c|}{\textbf{Matter era}} & $9.3 \cdot 10^{11}$ & $6.2 \cdot 10^{47}$ \\
   \cline{1-3}
   \multicolumn{1}{c|}{\textbf{Dark energy era}} & $3.7 \cdot 10^{17}$ & $5.4 \cdot 10^{58}$ \\
   \cline{1-3}
\end{tabular}
\end{center}
\end{table}

\begin{figure}[h!]
  \centering
  \includegraphics[scale=0.34]{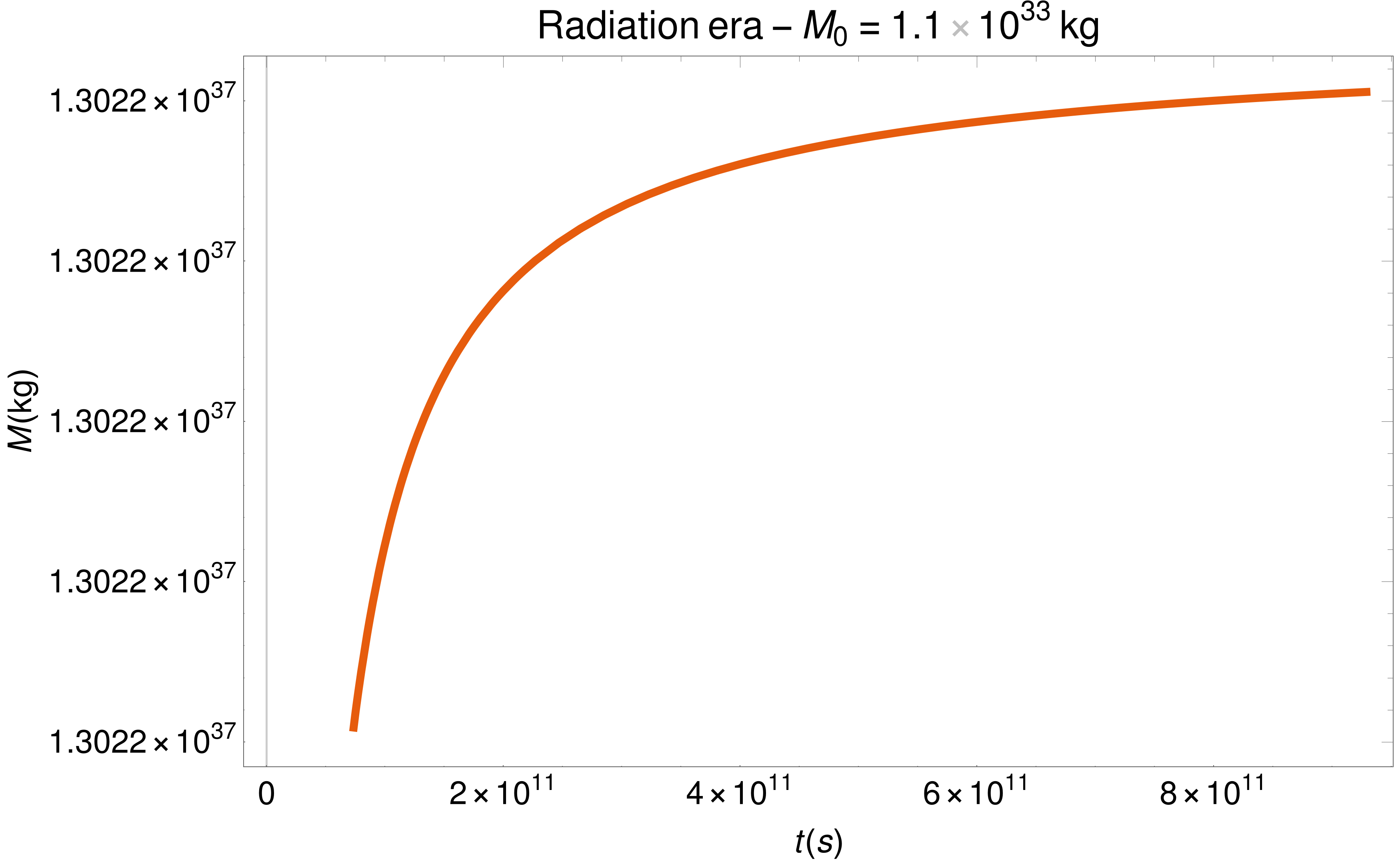}\\
  \includegraphics[scale=0.34]{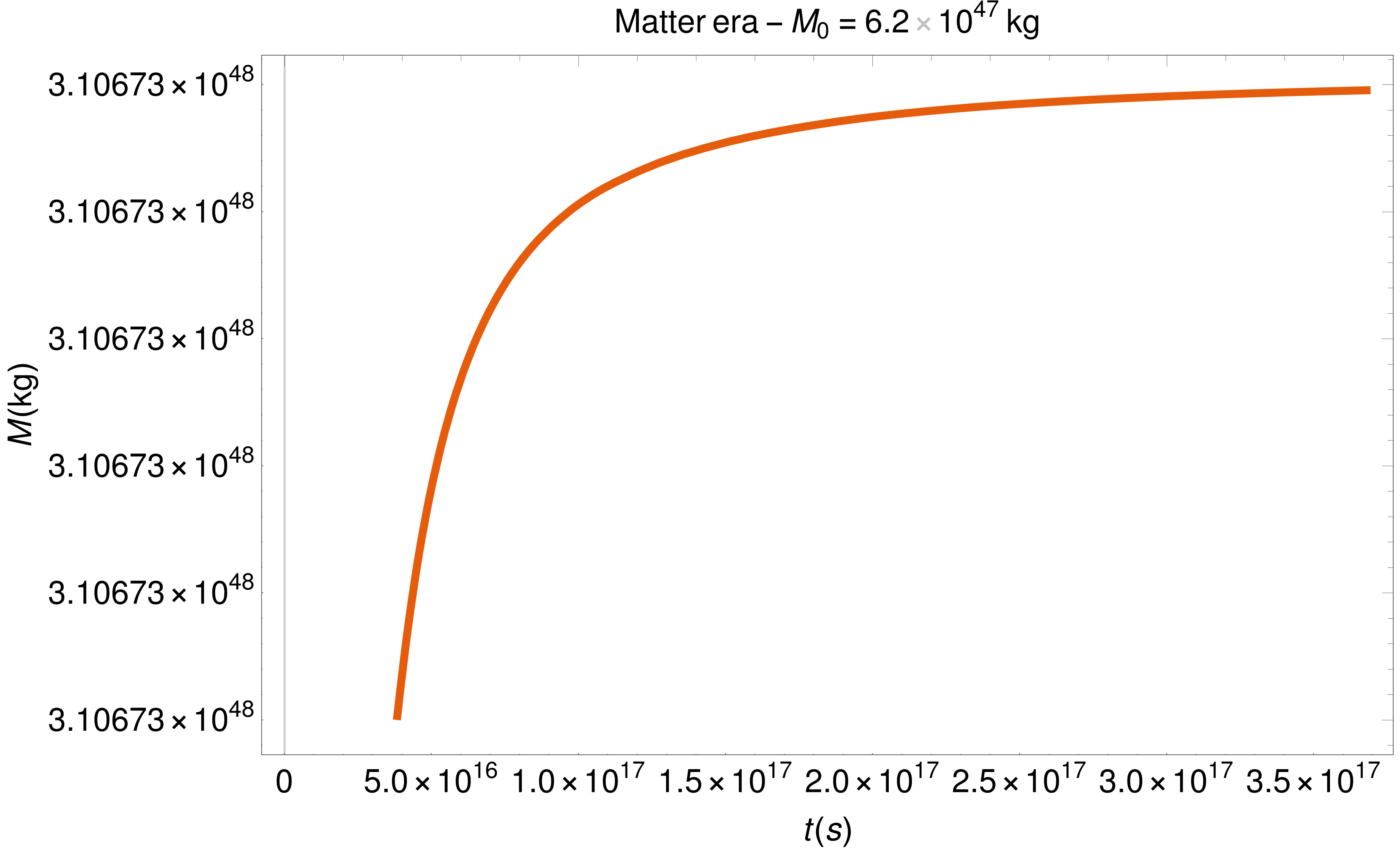}\\
  \includegraphics[scale=0.34]{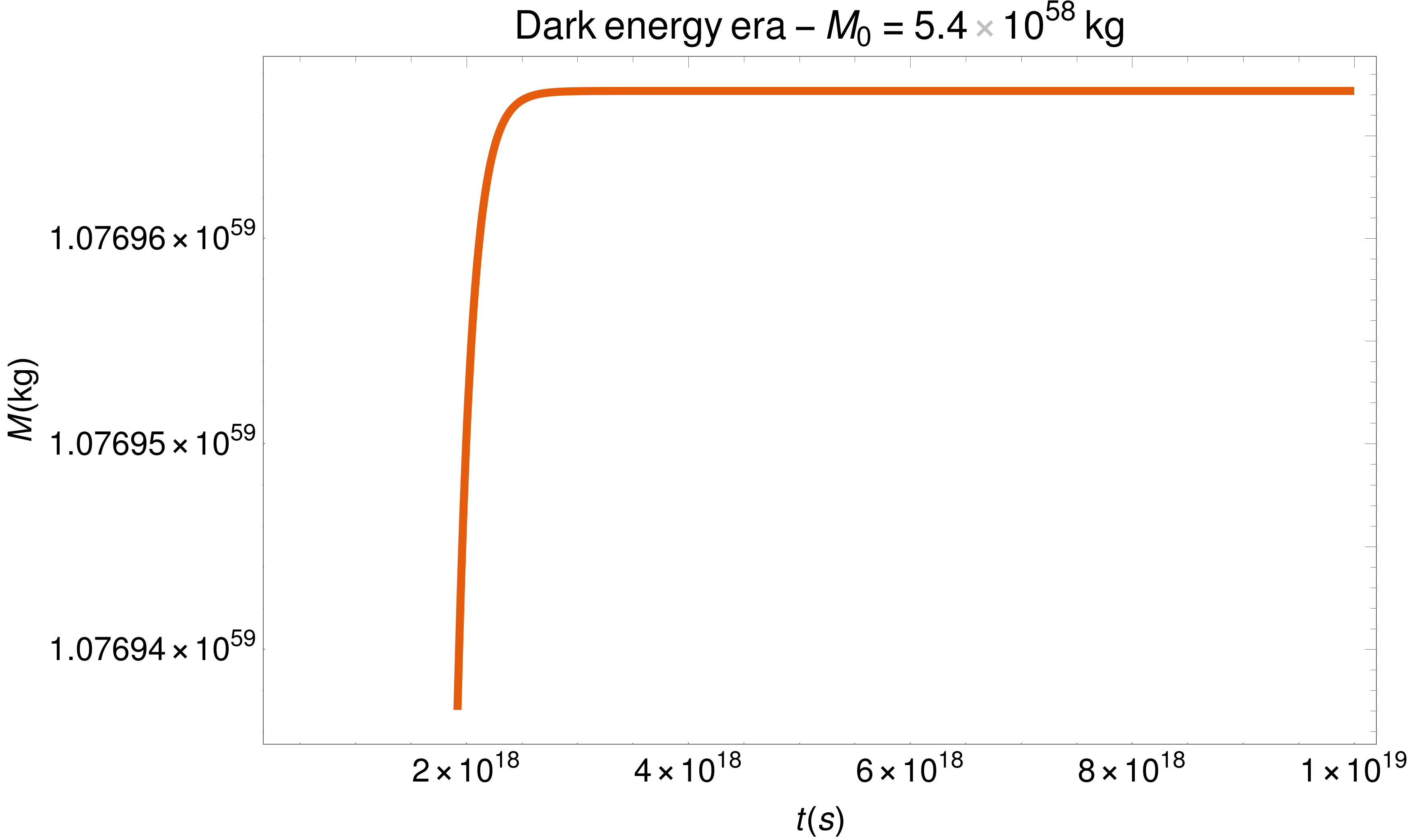}
  \caption{\label{fig:crescCRIT} Complete numerical solution for the mass evolution of a black hole immersed in cosmic microwave background radiation with temperature evolving according to the universe eras with initial masses that gives initially increasing behavior for each universe era (given by Table \ref{tab:mmincres}).}
\end{figure}

\begin{figure}[h!]
  \centering
  \includegraphics[scale=0.34]{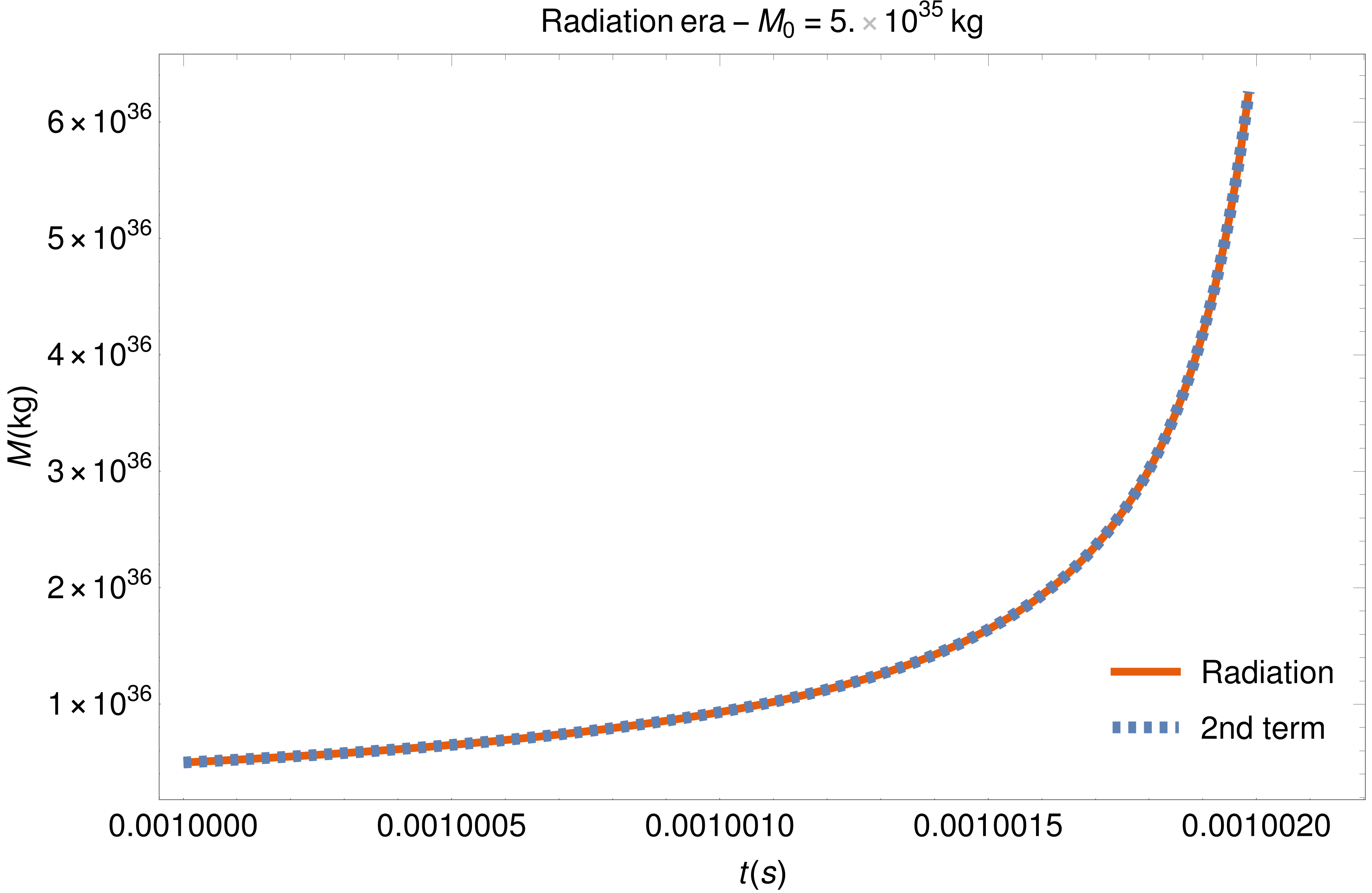}\\
  \includegraphics[scale=0.34]{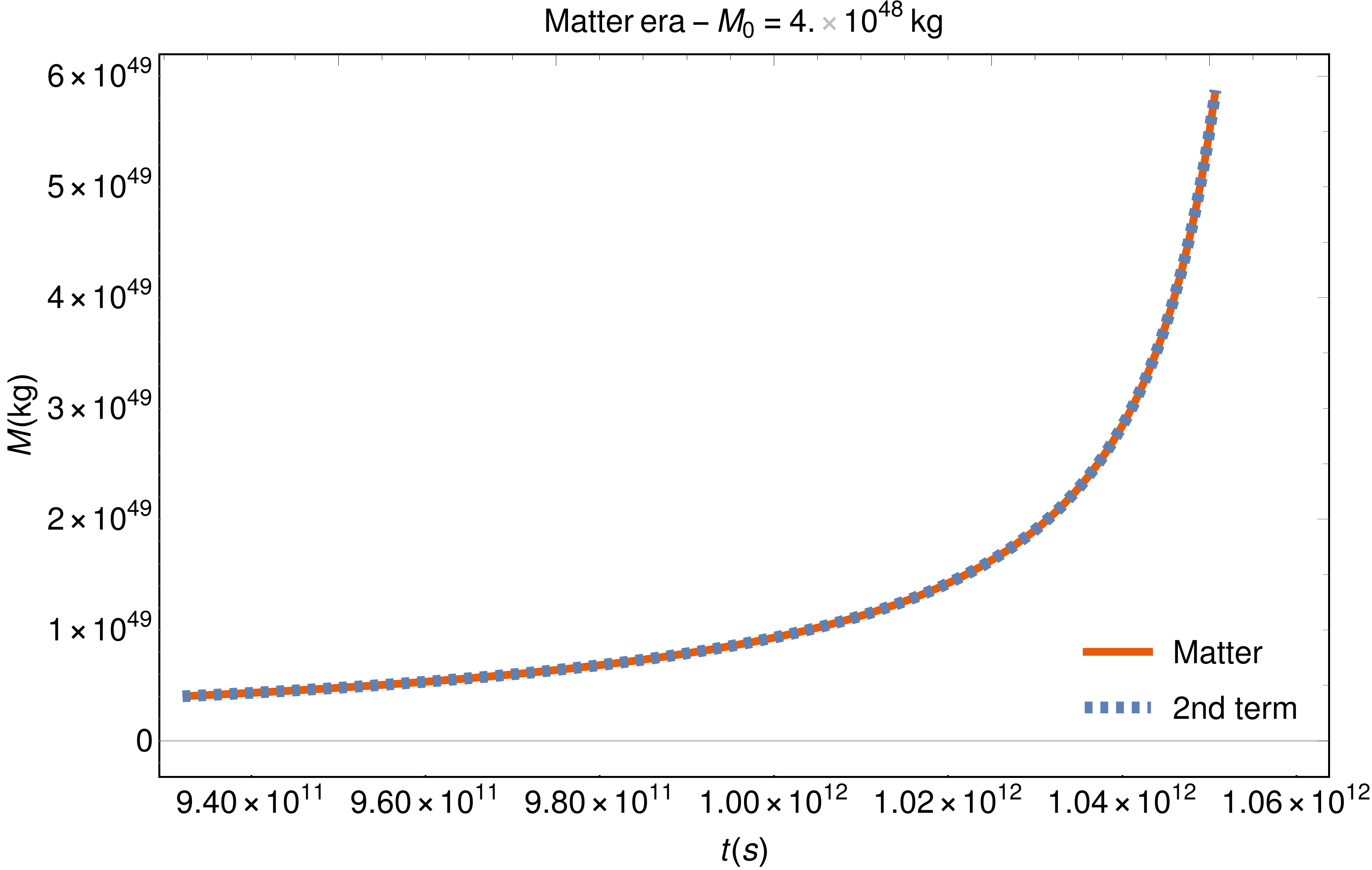}\\
  \includegraphics[scale=0.34]{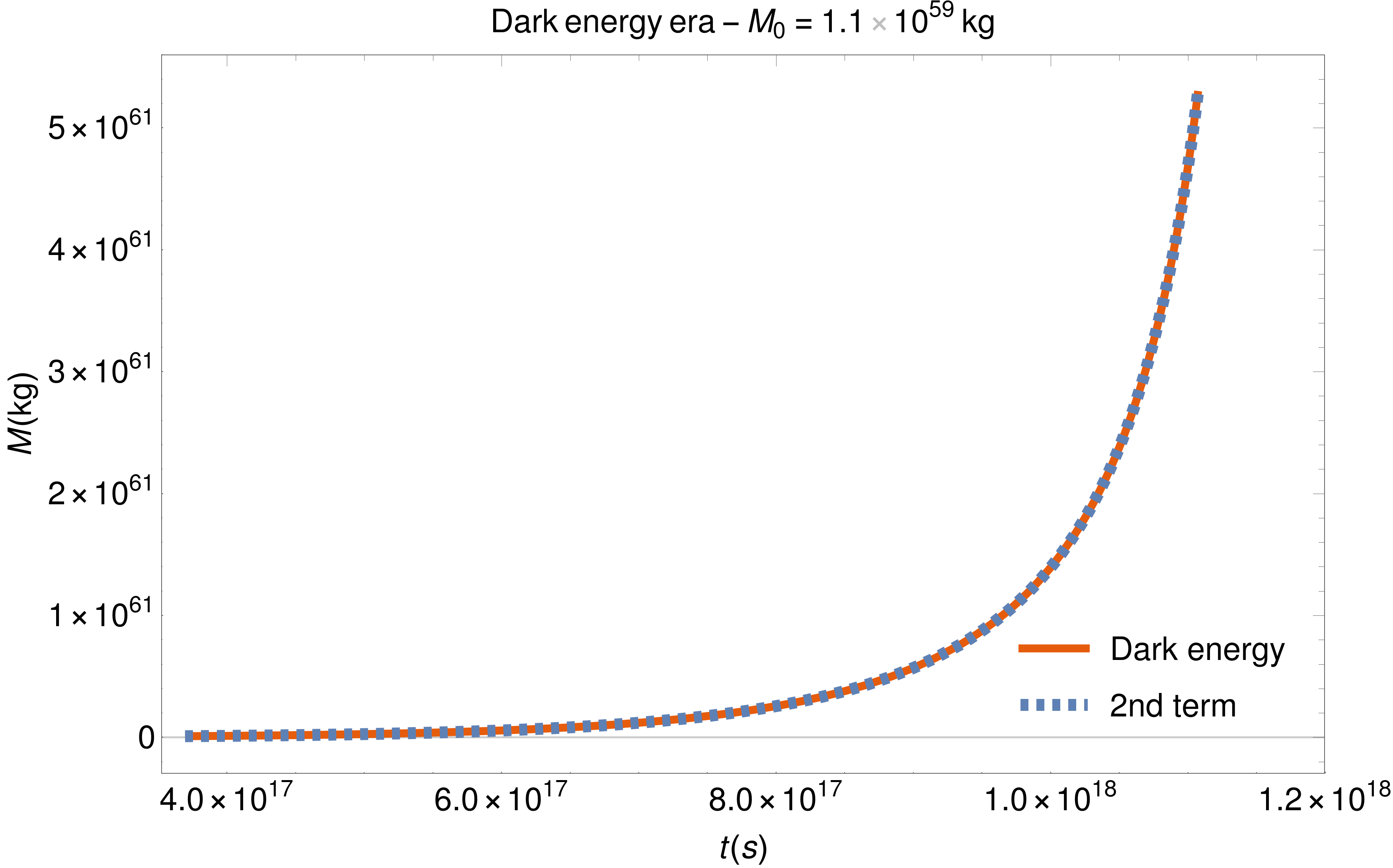}
  \caption{\label{fig:cresc} Solutions for the mass evolution of a black hole immersed in cosmic microwave background radiation with temperature evolving according to universe eras with initial masses greater than the ones given by Table \ref{tab:mmincres}. The plots show the analytical solution of the {\em 2nd term} and {\em complete numerical solutions} to each era, respectively. Notice that the curves do not reach the end of the eras, because the masses diverges before.}
\end{figure}

This evaluation is an estimate, because this analysis provides the behavior inversion (increasing instead of decreasing mass) for the plots in the {\em initial point} of each era and not along all points in question. We can see, for the initial conditions of Table \ref{tab:mmincres}, the increasing behavior of the mass for each era in the Figure \ref{fig:crescCRIT}.

Finally, for initial masses greater than those previously found with the anterior estimate, we expect the indefinite increasing of the black hole masses, with radiation absorption and decreasing of temperature, that can be seen in the Figure \ref{fig:cresc}.

In the plots of the Figure \ref{fig:cresc} we see the comparison of the {\em complete numerical solution} for each era with the asymptotic solution of the {\em 2nd term} of equations (\ref{eq:2rad}-\ref{eq:2dark}). In all the cases we can see the superposition of solutions, concluding that, one more time, the asymptotic solutions of the {\em 2nd term} describes with mastery the \emph{numerical solution} for the appropriate initial conditions. Besides that, the maximum time achieved by both curves, in the three plots, asymptotes the ones expected by the maximum value of the valid interval, (\ref{eq:int2rad}-\ref{eq:int2dark}), in the solutions of {\em 2nd term} (\ref{eq:2rad}-\ref{eq:2dark}). They are, respectively, $t_C^R \simeq 0.00100216 s$, $t_C^M \simeq 1.06103 \cdot 10^{12} s$ and $t_C^{\Lambda} \simeq 1.18256 \cdot 10^{18} s$. 

%---------------------------------------------------------------------

\section{Final considerations}
\label{sec:disc}

Until now, the possibility of a complete evaporation of a black hole is an open question. The point is that, when a black hole evaporates, it loses its event horizon, the entity that hides its inner states. Thus, all of the information inside the black hole disappears with its evaporation. This leads to what is called {\em information loss}. It occurs because, as suggested by semi-classical arguments, in the process of black hole formation and evaporation, a pure quantum state will evolve to a mixed state. Some possible alternatives to a non-existence of a complete evaporation of a black hole could be that the black hole evaporates until its length reaches the Planck scale, or the information comes out in a final burst, or that black holes quantum tunnels into a white hole, or even that no black hole ever forms. These ideas are discussed in the references \cite{unruh2017, ambrosio2018} but they are not explored in the present paper, because we have considered complete evaporation of the black holes, in their appropriate regimes.

There is another feature that we have not considered in the present paper: the ``origins'' of the black holes that we are working with, i.e., we have considered just the intervals for the masses, and not the formation process for the black holes; in special, we have not discussed if they are primordial or not. Besides that, we have also only considered the semi-classical approach, the Hawking radiation to the energy exchanges between the black hole and the environment. The reference \cite{horvath2005} considers the black hole evolution in the cosmological context, reviewing results related to thermodynamics of semi-classical black holes with influx of ambient particles and flux from its evaporation. The reference \cite{chirenti2007} discuss about evaporation of mini black holes.
And a more recent work \cite{nomura2019} presents a picture of the quantum mechanics of a collapse-formed, evaporating black hole. This one describes the radiation from the black hole in terms of ``hard modes'' and ``soft modes'', showing an entanglement structure between them and the Hawking radiation. Again, these considerations are out of the scope of this present work, but can be used to complement the analysis made by us.

It is interesting to make a discussion about the divergence of the masses of the black holes that is observed for some values of the parameters. Those situations happen in sections \ref{sec:rad} and \ref{sec:ETU}. In section \ref{sec:rad}, we have the black hole immersed in cosmic microwave background with constant temperature, so we expect $M \rightarrow \infty$ for black holes with $M_0 > M_{CMB}$ \eqref{eq:Mrcf}. In section \ref{sec:ETU} we still have the black hole immersed in cosmic microwave background, but with CMB temperature varying in time according to the universe eras. In both cases it is important to explain two main points. First, neither the complete numerical solutions, neither the asymptotic analytical solutions are able to provide the behavior of masses greater than the last pair of points in the respective plots (see Figure \ref{fig:EX2Tcte} in section \ref{sec:rad} and Figure \ref{fig:cresc} in section \ref{sec:ETU}) - numerically, the software used (\textit{Mathematica}) do not allow us to see bigger time intervals.

The second point is that the analytical asymptotic solutions ((\ref{eq:Msegundotermob}) for section \ref{sec:rad} and (\ref{eq:2rad}-\ref{eq:2dark}) for section \ref{sec:ETU}) have a finite time interval, i.e., there are critical times $t_C = \frac{1}{b M_0}$\footnote{See valid interval for (\ref{eq:Msegundotermob}).} and $t_C^i$ (for $i = R, M$ or $\Lambda$)\footnote{See valid intervals for (\ref{eq:int2rad}-\ref{eq:crescinterasparou}).} where the black holes masses diverge in a finite time. For time-values greater than the critical time, the masses become negative, and our equations are no longer valid (physically no-reasonable region). In this way, for $t > t_C$, our model loses its valid. Moreover, physically we can think that, while we increase the black hole mass, using a finite number of physical processes, we decrease its temperature and, consequently, its surface gravity. Therefore, the {\em 3rd law of black hole thermodynamics} arises in this limit and prevents the masses to diverge in a finite time, implying that some other physical effects must take place to prevent the divergence of the mass. Besides all that, when the mass goes to infinity, so does the Schwarzschild radius (the event horizon radius of the black hole), meaning that the black hole is getting arbitrarily big, which is clearly not realistic, since it would interact with other fields and objects, and our assumptions are no longer valid (neither our equations).

In the case of a black hole immersed in cosmic microwave background with temperature evolving in time we have the division of time intervals in eras. So, we have the same black hole with different initial masses at the beginning of each era, where its final mass in one era is equal to its initial mass in the subsequent era. Therefore, the behavior of the black hole (increase or decrease of its mass) can change between the eras. If we observe the table \ref{tab:mmincres} we can say that, for the black hole to increases its mass in the radiation era, the same has to have an initial mass of $M_{0_R} \gtrsim 1.1 \cdot 10^{33} kg$, but the mass cannot be much greater than this value, because, if we observe the figure \ref{fig:cresc}, it will diverge for $M_{0_R} \sim 5 \cdot 10^{35} kg$. Hence, the allowed interval for the initial mass which causes the mass to increase throughout the radiation era is $1.1 \cdot 10^{33} \lesssim M_{0_R} \lesssim 5 \cdot 10^{35} kg$. The same interval for the matter and dark energy eras are $6.2 \cdot 10^{47} \lesssim M_{0_M} \lesssim 4 \cdot 10^{48} kg$ and $5.4 \cdot 10^{58} \lesssim M_{0_\Lambda} \lesssim 1.1 \cdot 10^{59} kg$ respectively.

Notice that these intervals are very narrow, which means that our equations are very sensitive about great values of masses. In order for the masses to grow, but not diverge, throughout each era, the initial masses must satisfy these intervals. In special, if a black hole possess initially $1.1 \cdot 10^{33} \lesssim M_{0_R} \lesssim 5 \cdot 10^{35} kg$ in the radiation era, it is possible to reach the matter era with $6.2 \cdot 10^{47} \lesssim M_{0_M} \lesssim 4 \cdot 10^{48} kg$, and then reach the dark energy era with $5.4 \cdot 10^{58} \lesssim M_{0_\Lambda} \lesssim 1.1 \cdot 10^{59} kg$. In this case, the mass will always increase, from the beginning of the universe until today.

This paper contains a broad discussion about the mass evolution of Schwarzschild black holes. We have considered the black holes immersed in different environments like empty space and the cosmic microwave background (with constant and evolving in time temperature), and in all cases we considered the Stefan-Boltzmann law to model the evolution of the masses in time. We have analyzed all the equations according to asymptotic solutions and compared the numerical results to them, achieving the same results in all solutions and all cases. Besides that, we have considered several initial mass conditions to illustrate the different behaviors of black holes, with masses increasing or decreasing.

Finally, it is worthy to say that this paper summarizes the Master Degree dissertation of Natali Soler Matubaro de Santi, that can be viewed, in Portuguese, for more details, in \cite{natali2018}.

%---------------------------------------------------------------------

\begin{acknowledgements}

We would like to thank Prof. Dr. Alberto Vazquez Saa for overall support and comments. We are also grateful to ``Conselho Nacional de Desenvolvimento Cientifico e Tecnologico'' (CNPq) for financial support.

\end{acknowledgements}

%------------------------------------------------------------------------
%	APPENDIX
%------------------------------------------------------------------------

\vspace{0.3cm}
\begin{center}
{\large APPENDIX}
\end{center}
\vspace{-0.7cm}
  
\appendix

\section{Thermal evolution of the universe}
\label{sec:etu}

The thermal evolution of the universe can be viewed using the flat universe model ($\varkappa = 0$ in the Friedmann's equations (\ref{eq:F1}) and (\ref{eq:F2})) \cite{carroll2004, hartle2003, thorne2017}. In the present appendix we are going to introduce the Friedmann-Lemaître-Robertson-Walker (FLRW) metric, Friedmann's equations, energy momentum tensor for a perfect fluid and the necessary approximations to this model.

We know that the universe is homogeneous, isotropic and evolves in time. To describe the universe we use the FLRW metric
\begin{equation}
  \label{eq:friedmann}
  ds^2 = - dt^2 + a(t)^2 \left[ \frac{dr^2}{\left(1 - \varkappa r^2\right)} + r^2 \left( d \theta^2 + \sin^2 \theta \hspace{0.05cm}d \phi^2 \right)\right] ,
\end{equation}
where $a (t)$ is the expansion factor of the universe, $r$ is the radial coordinate, $\theta$ and $\phi$ are the angular spherical coordinates and $\varkappa$\footnote{The curvature parameter $\varkappa$ classifies the universe as: \emph{closed} with positive curvature ($\varkappa = 1$), \emph{plane} and without curvature ($\varkappa = 0$) and \emph{opened} with negative curvature ($\varkappa = - 1$).} a parameter related with the curvature and with dimension $L^{-2}$. The perfect fluid consideration\footnote{We choose the 4-velocities being $u^{\mu} = (1, 0, 0, 0)$.} as curvature generator and the Einstein's equations lead us to the so-called \emph{Friedmann's equations}, and these equations define the time evolution of matter and energy. They can be written as
\begin{align}
 \left(\frac{\dot{a}}{a}\right)^2 & = \frac{8 \pi}{3} \rho - \frac{\varkappa}{a^2} , \label{eq:F1}\\
 \frac{\ddot{a}}{a} & = - \frac{4 \pi}{3} \left( \rho + 3 p \right) , \label{eq:F2}
\end{align}
with $\dot{a} = \frac{d a}{d t}$, $\rho$ the matter density and $p$ its pressure. Besides Friedmann's equations, we can use some properties of the energy-momentum tensor and compute
\begin{equation}
  \nabla_{\mu} T^{\mu}_{0} = \partial_0 (- \rho) + \left(\Gamma^1_{0 1} + \Gamma^2_{0 2} + \Gamma^3_{0 3}\right)(- \rho) - \left(\Gamma^1_{0 1} + \Gamma^2_{0 2} + \Gamma^3_{0 3}\right) p .
\end{equation}
Since $\nabla_{\mu} T^{\mu \nu} = 0$ we have
\begin{eqnarray}
  - \partial_t \rho - \frac{3}{2} \frac{\partial_t a^2}{a^2} \rho = \frac{3}{2} \frac{\partial_t a^2}{a^2} p .
\end{eqnarray}
Note that, writing $\partial_t a^3 = 3 a^2 \partial_t a$ and $\partial_t a^2 = 2 a \partial_t a$, we get the following \emph{state equation}
\begin{equation}
  \frac{d (a^3 \rho)}{d t} = - p \frac{d a^3}{d t} . \label{eq:est}
\end{equation}
It is worthy to say that this equation (\ref{eq:est}) does not replace Friedmann's equations because, although it provides a temporal evolution of the matter and energy content of the universe, it is not straightly related to the universe curvature.

\begin{figure}[h!]
  \centering
  \includegraphics[scale=0.29]{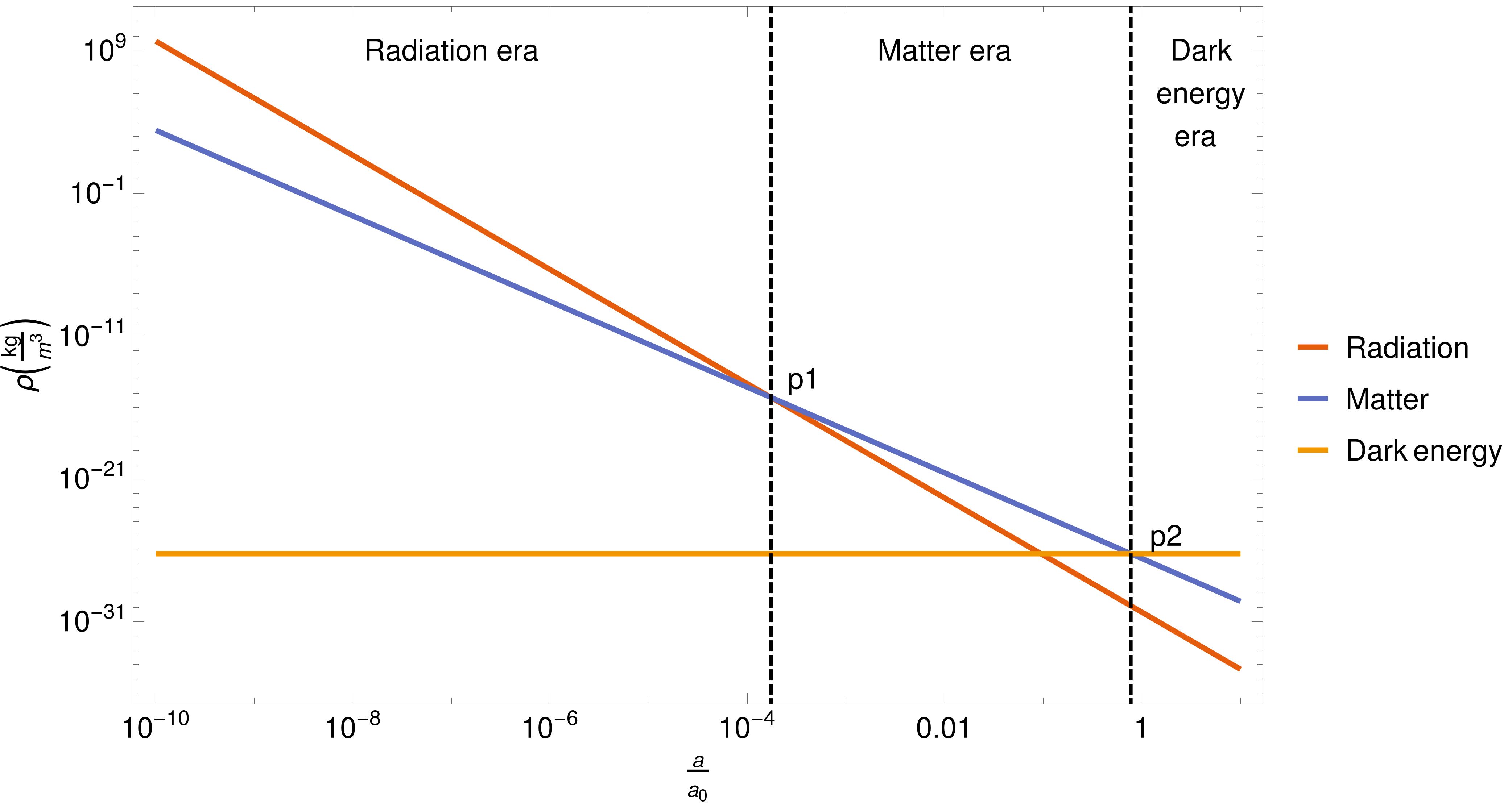}
  \caption{\label{fig:eras} Density evolution of radiation, matter and dark energy according to expansion factor $\frac{a}{a_0}$ of the universe. Note that we have used the logarithm scale in both axes.}
\end{figure}  

The division of the matter and energy content of the universe is: \emph{radiation}, \emph{matter} and \emph{dark energy}. The \emph{radiation} includes the photons, the cosmic microwave background, the gravitons and the neutrinos, i.e., particles that satisfies the state equation $p_R = \frac{\rho_R}{3}$. By \emph{matter} we refer to hadronic and dark matter\footnote{The dark matter means some kind of unhadronic matter that does not interact with electromagnetic radiation.}, entities whose pressure will be negligible when compared to the density: $p_M = 0$. And finally, the \emph{dark energy}\footnote{By dark energy we mean some kind of energy that is still unknown and extends uniformly throughout all the universe and seems like ``vacuum energy''. Because of it, one of its proposals is related to non-null \emph{cosmological constant} $\Lambda$.} satisfies $p_{\Lambda} = - \rho_{\Lambda}$. Given those state equations we can integrate the equation (\ref{eq:est}) and obtain
\begin{align}
  \rho_R & = \rho_{R_0} \left(\frac{a_0}{a}\right)^4 \label{eq:rhoR}\\
  \rho_M & = \rho_{M_0} \left(\frac{a_0}{a}\right)^3 \label{eq:rhoM}\\
  \rho_{\Lambda} & = \rho_{\Lambda_0} , \label{eq:rhoD}
\end{align}
where $\rho_{i_0}$ is the radiation density ($i = R$), matter density ($i = M$) and dark energy density ($i = \Lambda$) today and $a_0$ is the expansion factor of the universe today.

Using the experimental values, found in \cite{pdg2014}, to the radiation, matter and dark energy, we compute their respective present densities $\rho_{R_0} \simeq 4.6 \cdot 10^{- 31} \frac{kg}{m^3}$, $\rho_{M_0} \simeq 2.7 \cdot 10^{- 27} \frac{kg}{m^3}$ and $\rho_{\Lambda_0} \simeq 5.8 \cdot 10^{- 27} \frac{kg}{m^3}$ and we plot the evolution of the densities $\rho_{R}$, $\rho_{M}$ and $\rho_{\Lambda}$ \textit{versus} the expansion factor of the universe normalized $\frac{a}{a_0}$, viewed in the Figure \ref{fig:eras}. Observe that we have highlighted three different phases, that we called of \emph{radiation era}, \emph{matter era} and \emph{dark energy era}. Every phases is dominated by one matter and energy content of the universe. They are the so-called \textit{universe eras}. The divisions are represented by the vertical dashed lines, where the densities intersect each other: $p_1 = \frac{a}{a_0} = \frac{\rho_{R_0}}{\rho_{M_0}} \simeq 1.7 \cdot 10^{-4}$ and $p_2 = \frac{a}{a_0} = \left(\frac{\rho_{M_0}}{\rho_{\Lambda_0}}\right)^{1/3} = 7.7 \cdot 10^{-1}$.

Still using the experimental values of the Reference \cite{pdg2014}, we can say that nowadays the universe is flat, in other words, the curvature parameter $\varkappa = 0$. This statement can be viewed rewriting equation (\ref{eq:F1}) as
\begin{equation}
 \Omega - 1 = \frac{\varkappa}{H^2 a^2} , \label{eq:reF1}
\end{equation}
where $\Omega = \frac{8 \pi}{3 H^2} \rho = \frac{\rho}{\rho_{crit}}$ is the \emph{density parameter}, $\rho_{crit} = \frac{3 H^2}{8 \pi}$ is the \emph{critical density} and $H = \frac{\dot{a}}{a}$ the \emph{Hubble's parameter}. As $\Omega$ is the sum of the parameters of radiation, matter and dark energy densities, using their experimental values we can see that
\begin{equation}
 \Omega = \Omega_{R} + \Omega_{M} + \Omega_{\Lambda} \simeq 5.46 \cdot 10^{-5} + 0.315 + 0.685 \sim 1 ,
\end{equation}  
which implies, by equation (\ref{eq:reF1}), that $\varkappa = 0$.

Now we are going to connect the time evolution of the matter and energy content of the universe with Einstein's equations. To do this, we just need to replace the results (\ref{eq:rhoR}-\ref{eq:rhoD}) in the Friedmann's equations (\ref{eq:F1}) and (\ref{eq:F2}). In order to do this, let's then consider a flat universe ($\varkappa = 0$) not just today but throughout the evolution of the universe (as commented above)\footnote{This choice is our main approximation to this model.}. Thus, replacing the densities in the $1st$ Friedmann's equations (\ref{eq:F1}), we find
\begin{align}
 \left(\frac{a}{a_0}\right) & = \left(\frac{32 \pi}{3} \rho_{R_0}\right)^{1/4} t^{1/2} , \hspace{0.1cm}t \in (0, t_{p_1})\label{eq:aa0rad}\\
 \left(\frac{a}{a_0}\right) & = (6 \pi \rho_{M_0})^{1/3} t^{2/3} , \hspace{0.1cm} t \in (t_{p_1}, t_{p_2})\label{eq:aa0mat}\\
 \left(\frac{a}{a_0}\right) & = \textnormal{Exp}\left[\left(\frac{8 \pi}{3} \rho_{\Lambda_0} \right)^{1/2} t\right] , \hspace{0.1cm}t \in (t_{p_2}, \infty), \label{eq:aa0dark}
\end{align}
where, recovering the units using SI, $t_{p_1} = \left[\frac{3}{32 \pi G \rho_{R_0}} \left(\frac{\rho_{R_0}}{\rho_{M_0}}\right)^4\right]^{1/2} \simeq 9.3 \cdot 10^{11} s$ and $t_{p_2} = \frac{1}{(6 \pi G \rho_{\Lambda_0})^{1/2}} \simeq 3.7 \cdot 10^{17} s$, and these values were obtained replacing $p_1$ and $p_2$ respectively in (\ref{eq:aa0rad}) and (\ref{eq:aa0mat}). Note that we integrate indefinitely, in $a$ and in $t$, the expressions (\ref{eq:aa0rad}) and (\ref{eq:aa0mat}), disregarding any integration constant. We have also integrated indefinitely in (\ref{eq:aa0dark}), defining $a_0$ as integration constant. This is an approximation to the solution of (\ref{eq:F1}) in terms of dominance intervals of radiation, matter and dark energy, and the result stays undefined due to a lack of the integration constants. However, since we are not interested in the continuity of the solution to equation (\ref{eq:F1}), but in its form, our result is convenient to our analysis.

Finally, we need to relate the temperature with the universe densities. We know that a black body emits radiation with energy density according to Stefan-Boltzmann's law, i.e., $L \propto T^4$. We saw that the density radiation of the universe can be written as $\rho_R \propto \left(\frac{a_0}{a}\right)^4$. Thus, doing a dimensional analogy, we can say that the radiation temperature evolves as 
\begin{equation}
 T_{R} = T_{CMB} \left(\frac{a_0}{a}\right) , \label{eq:Tr}
\end{equation}
where $T_{CMB}$ is the temperature of the cosmic microwave background today. Therefore, considering that the radiation has temperature $T_i$, with $i = R, M, \Lambda$, related to the fraction $\frac{a_0}{a}$, and that this fraction assumes different expressions according to the universe eras: radiation ($R$), matter ($M$) and dark energy ($\Lambda$), we have
\begin{align}
 T_R & = \frac{T_{CMB}}{\left(\frac{32 \pi}{3} \rho_{R_0}\right)^{1/4}} \frac{1}{t^{1/2}}, \hspace{1cm} t \in (0, t_{p_1}) , \nonumber\\
 T_M & = \frac{T_{CMB}}{\left(6 \pi \rho_{M_0}\right)^{1/3}} \frac{1}{t^{2/3}} ,  \hspace{1cm} t \in (t_{p_1}, t_{p_2}) , \label{eq:Traderas}\\
 T_{\Lambda} & = T_{CMB} \hspace{0.1cm} \textnormal{Exp} \left[- \left(\frac{8 \pi}{3} \rho_{\Lambda_0}\right)^{1/2} t\right] , \hspace{1cm} t \in (t_{p_2}, \infty) . \nonumber
\end{align}

Hence, we can use these results in subsection \ref{sec:ETU} assuming that the black hole is exchanging energy with the universe, considering the three different ways of thermal evolution.

%---------------------------------------------------------------

% Authors must disclose all relationships or interests that 
% could have direct or potential influence or impart bias on 
% the work: 
%
% \section*{Conflict of interest}
%
% The authors declare that they have no conflict of interest.

% BibTeX users please use one of
%\bibliographystyle{spbasic}      % basic style, author-year citations
%\bibliographystyle{spmpsci}      % mathematics and physical sciences
%\bibliographystyle{spphys}       % APS-like style for physics
%\bibliography{}   % name your BibTeX data base

% Non-BibTeX users please use

%----------------------------------------------------

\end{document}